\documentclass[aps,pre,twocolumn,twoside,amsmath,amssymb,aps,floatfix,showpacs]{revtex4-2}

\usepackage{xcolor}
\usepackage{graphicx}
\usepackage{subfigure}
\usepackage{dcolumn}
\usepackage{bm}
\usepackage{framed}
\usepackage{hyperref}
\usepackage{dsfont}
\usepackage[normalem]{ulem}
\usepackage[utf8]{inputenc}

\newcommand\mean[1]{\langle #1 \rangle}

\begin{document}
	
\title{Scaling of multicopy constructive interference of Gaussian states}

\author{Matthieu Arnhem}
\email[]{matthieu.arnhem@univ-lille.fr}
\affiliation{Univ. Lille, CNRS, Inria, UMR 8524 - Laboratoire Paul Painlevé, F-59000 Lille, France}
\affiliation{Department of Optics, Palacký University, 17. listopadu 1192/12, 77146 Olomouc, Czech Republic}
\author{Radim Filip}
\email[]{filip@optcs.upol.cz}
\affiliation{Department of Optics, Palacký University, 17. listopadu 1192/12, 77146 Olomouc, Czech Republic}

\begin{abstract}
	Quantum technology advances crucially depend on the scaling up of essential quantum resources. Their ideal multiplexing offers more significant gains in applications; however, the scaling of the nonidentical, fragile and {varying} resources is neither theoretically nor experimentally known. For bosonic systems, multimode interference is an essential tool already widely exploited to develop quantum technology. Here, we analyze, predict and compare essential scaling laws for  {a constructive} interference of multiplexed nonclassical Gaussian states carrying information by displacement with weakly fluctuating squeezing in different multimode interference architectures.  {The signal-to-noise ratio quantifies the increase in displacement relative to the noise.} We introduce the \textit{gain-to-instability ratio} to numerically estimate the effect of unexplored resource instabilities in a large scale interference scheme. The use of the gain-to-instability ratio to quantify the scaling laws opens steps for extensive theoretical investigation of other bosonic resources and follow-up feasible experimental verification necessary for further development of these platforms.
\end{abstract}

\maketitle

\nopagebreak

\section{Introduction}\label{sec:intro}

\begin{figure*}[ht!]
	\centering
	\includegraphics[width=1.0\linewidth]{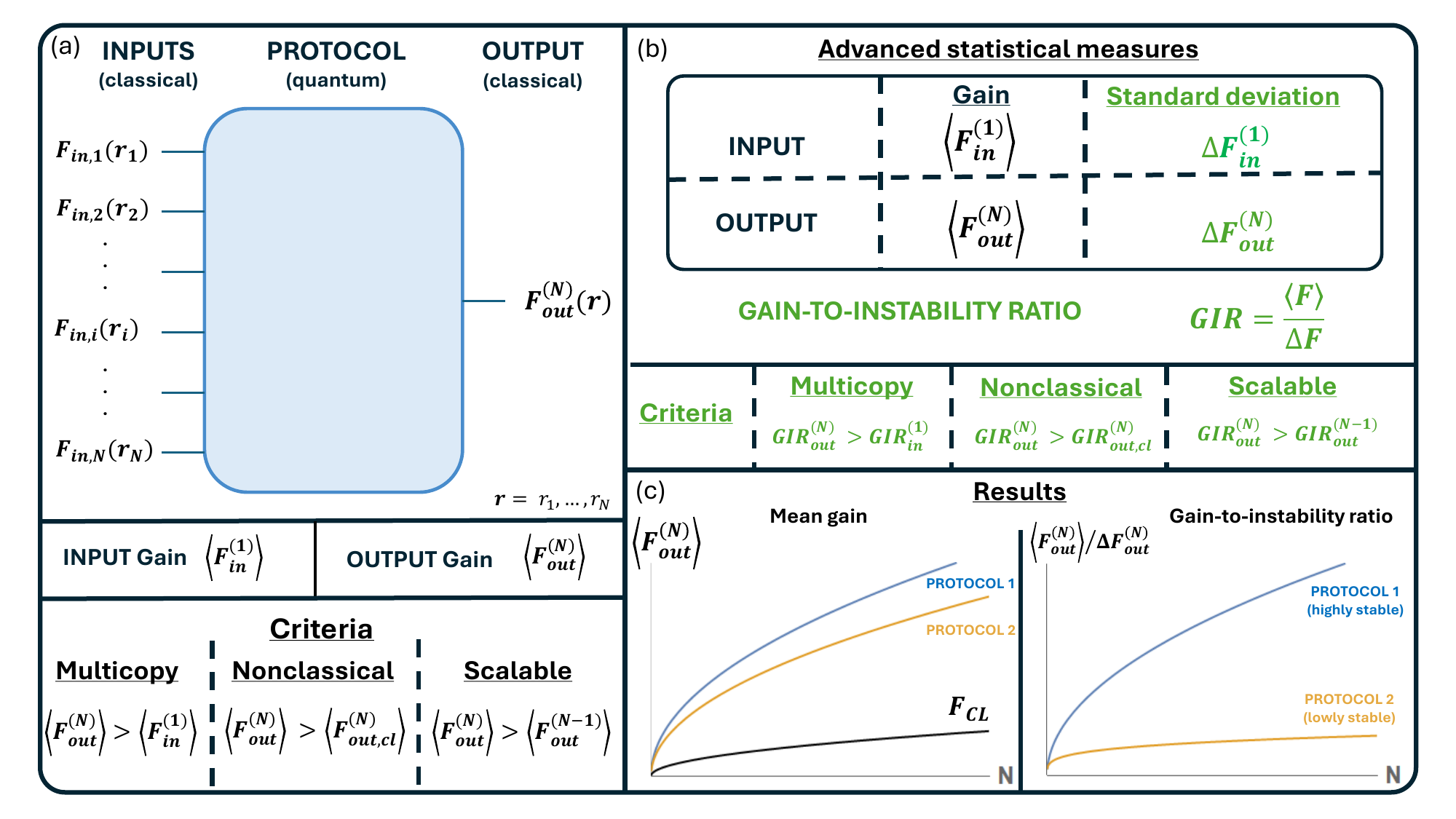}
\caption{  {Gain-to-instability quantification of quantum protocols:} (a) multicopy protocols and their applications use quantum resources parametrised by $\boldsymbol{r}= (r_1,...,r_N)$ in a number $N$ of finite,  {different} and imperfect replicas of inputs quantified by the classical figure-of-merit $F_{in,i}(r_i)$, $i=1,\ldots,N$. The  {quantum states carrying resources} are interacting in a quantum protocol that produces an output figure-of-merit $F_{out}^{(N)}(\boldsymbol{r})$ dependent on  {all} the resources $r_1,\ldots,r_N$.  Each input is characterized by the same  {gain as average figure of merit} $\langle F_{in}^{(1)}\rangle = \mean{F_{in,i}(r_i)}$ where $i=1,...,N$. 
 {If the gains $\mean{F^{(N)}_{out}} > \mean{F_{in}^{(1)}}$ than any of the input,} therefore it exhibits  {a multicopy} gain.  {The gain may be considered nonclassical when it surpasses the classical benchmark $\mean{F_{out, cl}^{(N)}}$ and scalable if it is larger than $\mean{F_{out, cl}^{(N-1)}}$, for any $N$.} (b)  {To thoroughly evaluate the protocol, we propose an additional quantifier, the \textit{gain-to-instability ratio} (GIR) for both inputs and outputs. We also extend the multicopy, nonclassical, and scalable criteria for GIR.}  As depicted in (c),  {the protocols with comparably scalable gains exhibit qualitatively different scaling of stability in the GIR. Protocol 1 (in blue) exhibits a highly stable scalability and, on the other hand, the protocol 2 (in orange) behaviour a lowly stable scalability.}}
\label{fig:Descriptionfigure}
\end{figure*}

Quantum physics and technology with bosonic systems intimately depend on the broad class of resource states, from primary Gaussian coherent and squeezed states  \cite{cerf_quantum_2007,weedbrook_gaussian_2012-1} to advanced non-Gaussian states {\cite{walschaers_non-gaussian_2021-1,rakhubovsky_quantum_2024,lachman_quantum_2022,huang_optical_2015,hacker_deterministic_2019,simon_experimental_2024}, }nonlinear phase states \cite{gottesman_encoding_2001,marek_general_2018} up to complex grid  quantum states \cite{gottesman_encoding_2001,bravyi_universal_2005} and  multimode graph states \cite{walschaers_tailoring_2018,walschaers_emergent_2023}. Nonclassical optical and atomic states play essential roles as the probes, codes, and critical ancillary states in all applications of bosonic systems, starting from quantum clocks \cite{lewis-swan_robust_2018,schulte_prospects_2020,pedrozo-penafiel_entanglement_2020,robinson_direct_2024} and metrology \cite{degen_quantum_2017,stokowski_integrated_2023,malia_distributed_2022,xia_entanglement-enhanced_2023,franke_quantum-enhanced_2023,nielsen_deterministic_2023}, communication \cite{madsen_continuous_2012,gehring_implementation_2015,jacobsen_complete_2018,kovalenko_frequency-multiplexed_2021,suleiman_40_2022}, quantum computational algoritms \cite{zhong_quantum_2020,wang_boson_nodate,zhong_phase-programmable_2021} and continuing up to the development of resource states {\cite{asavanant_generation_2019,larsen_deterministic_2019,konno_logical_2024,larsen_integrated_2025}}and gates \cite{miwa_exploring_2014,larsen_deterministic_2021,sakaguchi_nonlinear_2023} towards scalable, fast and universal quantum computers \cite{asavanant_optical_2022}. Simultaneously, quantum non-Gaussian states of trapped ions and superconducting circuits demonstrated their first usefulness in the sensing \cite{wolf_motional_2019,mccormick_quantum-enhanced_2019,wang_heisenberg-limited_2019,podhora_quantum_2022,deng_heisenberg-limited_2023,pan_realisation_2024} and quantum error correction \cite{fluhmann_encoding_2019,campagne-ibarcq_quantum_2020,eickbusch_fast_2022,de_neeve_error_2022,ni_beating_2023,sivak_real-time_2023}. 

In all such experiments, physical resources are not only naturally finite and imperfect.  {Moreover, their amount is never entirely identical across their copies and is typically not sufficiently stable, but varies across the copies.} For the protocols, such resource states are theoretically modelled  as realistic by considering finite energy and with the most dominant imperfection. However, these models broadly omit that they are different and  {not stable} both in the amount of the resource and imperfections over multiple replicas of the resource. This seriously limits a critical analysis of feasibility, extendibility and scalability analysis for currently developed multicopy protocols. In many above mentioned applications, we then use many different copies of such probes, codes, or  ancillary states.  Therefore, the overall performance depends on these differences  in the resource states, their distributions, the protocols used, and the final figure of merit we aim for. These fundamental and practical thoughts open new questions about unexplored quantum scaling phenomena considering even slightly different resource states we use for applications. Such scalings will navigate further protocol developments and therefore, it can cross-fertilize developing quantum technology. 

 {Essential and paramount nonclassical squeezed states can provide the first experimentally testable cases with different resources that are extensively scalable \cite{yokoyama_ultra-large-scale_2013,zhong_quantum_2020,arrazola_quantum_2021,enomoto_programmable_2021,tiedau_statistical_2021}. Squeezed states as resources are already broadly applicable for {quantum sensing {\cite{ganapathy_sensing_2023,xia_sensing_2023,stokowski_sensing_2023,guo_sensing_2020,nielsen_sensing_2023}}} and quantum communication {\cite{madsen_continuous_2012,gehring_implementation_2015,jacobsen_complete_2018,kovalenko_frequency-multiplexed_2021,suleiman_40_2022}} but are also used for quantum computation primitives like Gaussian boson sampling {\cite{wang_boson_nodate}} or for cluster state quantum computing \cite{yokoyama_ultra-large-scale_2013,chen_experimental_2014,cai_multimode_2017,asavanant_generation_2019}. The time multiplexing of squeezed states has recently been scaled up to $10^6$ modes \cite{yoshikawa_invited_2016}. It can allow experimental testing of the scaling laws with an unprecedented number of nonclassical states.} 

 {In this paper, we quantify how a figure of merit scales with the number of copies impacted by resource instability in the multicopy protocol, as illustrated in Fig.1(a). We consider the figure of merit, averaged over the instability of resources, as the protocol \textit{gain}. We propose a dimensionless \textit{gain-to-instability ratio} (GIR), which relates the gain to the standard deviation of the variation in the figure of merit, see Fig.1(b). We evaluate the increase in gain and GIR as a function of the number of copies. The GIR enables us to discuss scaling beyond the gain and to differentiate between unknown limitations of multicopy experiments, as demontrated in Fig.1(c).} 
	
 {By employing the GIR, we analyse the constructive interference of the displaced Gaussian states with {varying noise squeezing across the copies} in different interference architecture schemes and uncover their scalability. The figure of merit for our essential interference investigation is the signal-to-noise ratio (SNR). SNR variation caused by unstable squeezing over many copies limits the scaling of GIR. Note that in the ideal case where all the input states are {identical and perfectly stable} states, the GIR diverges to infinity. Hence, limits on the GIR is a signature of imperfect input state copies. The scalings of the gain and GIR are suitable for immediate experimental tests \cite{yonezu_time-domain_2023, enomoto_programmable_2021, takeda_-demand_2019} and facilitates further theoretical predictions for other Gaussian and quantum non-Gaussian resources used in bosonic systems and their applications.}

\begin{figure*}[!ht]
	\centering
	\begin{subfigure}{\includegraphics[width=0.31\textwidth]{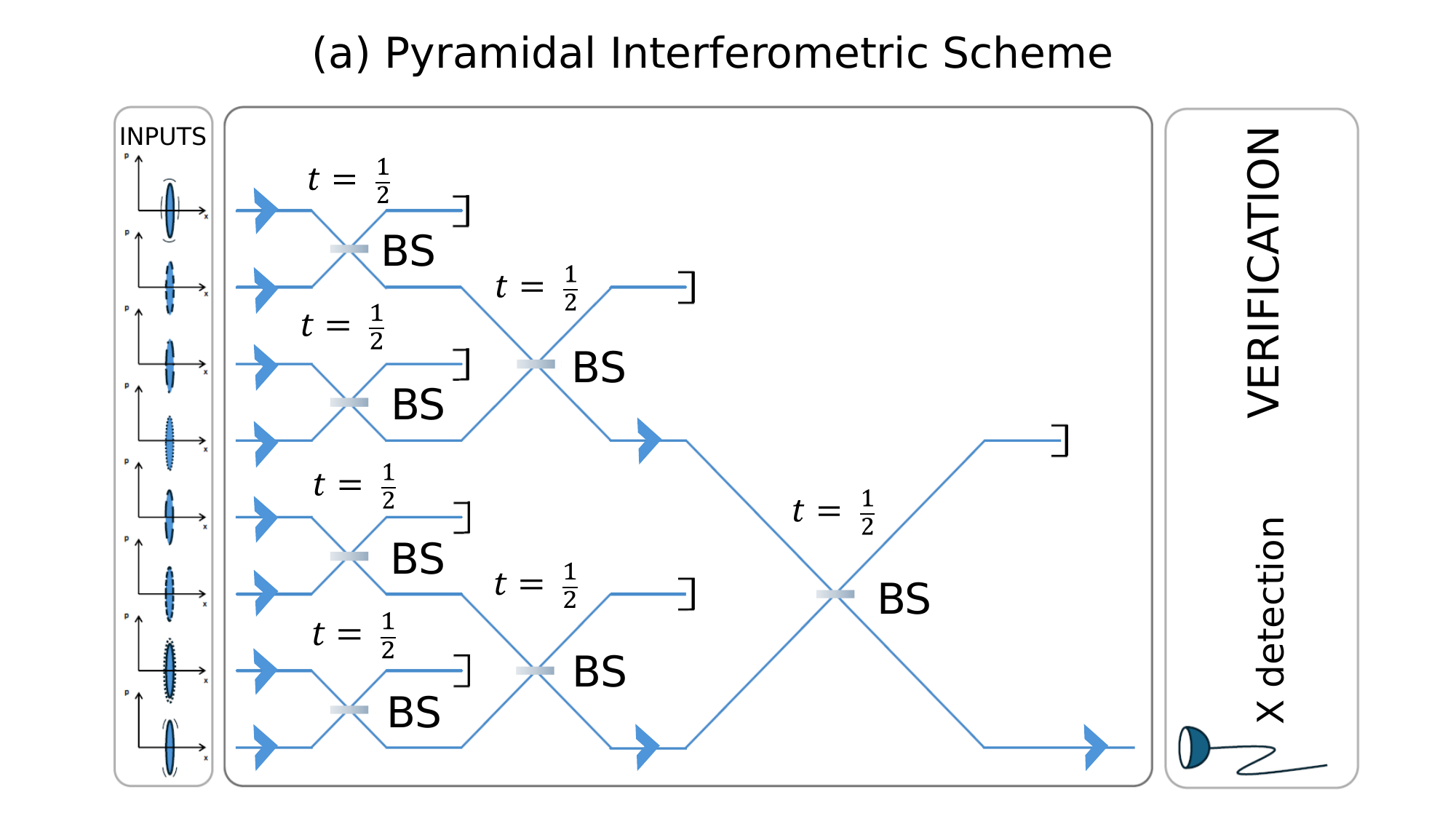} 
	\label{fig:pyramidalarchitecture}}
	\end{subfigure}
	\begin{subfigure}{\includegraphics[width=0.31\textwidth]{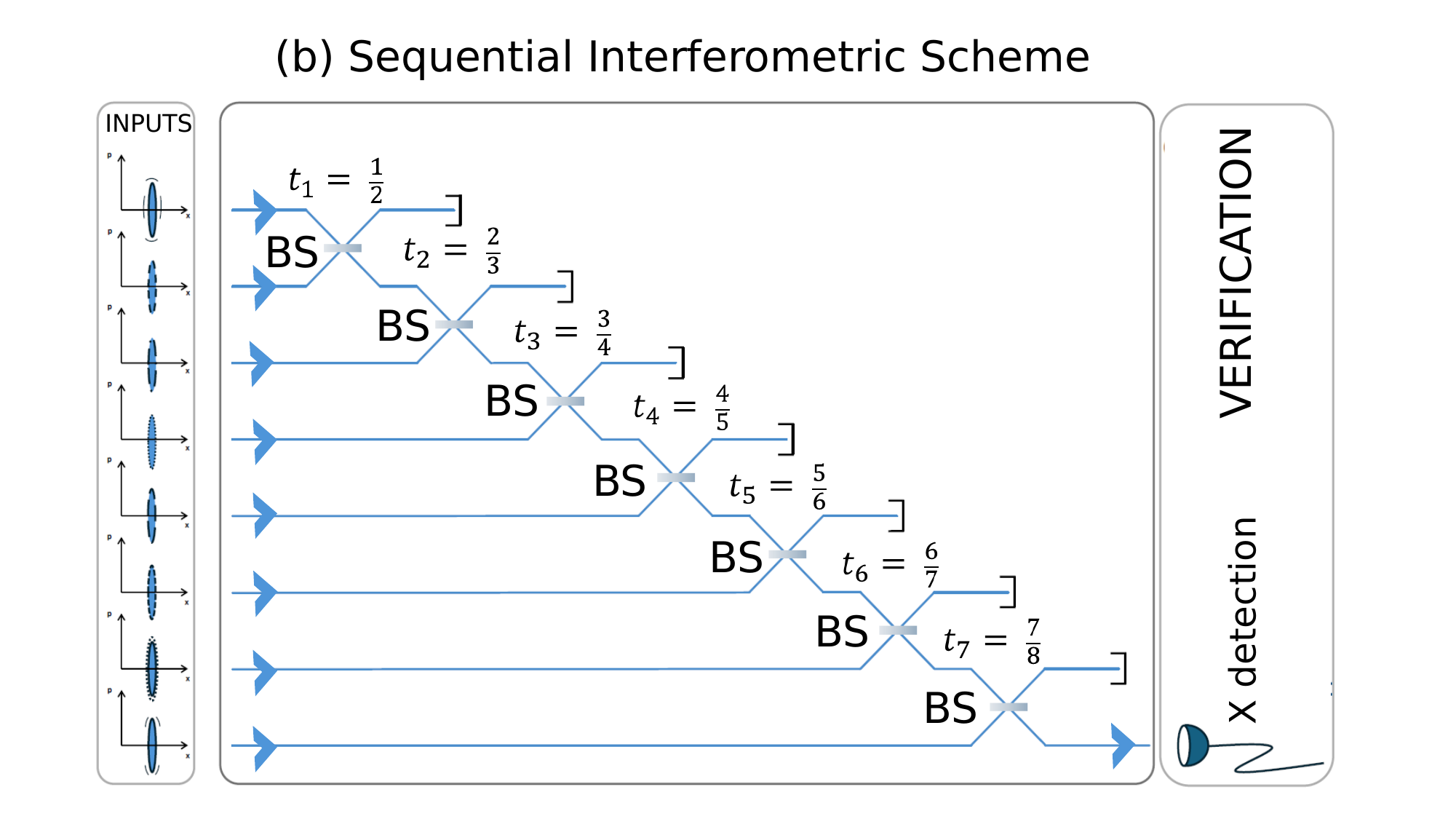}
	\label{fig:sequentialarchitecture}}
	\end{subfigure}
	\begin{subfigure}{\includegraphics[width=0.31\textwidth]{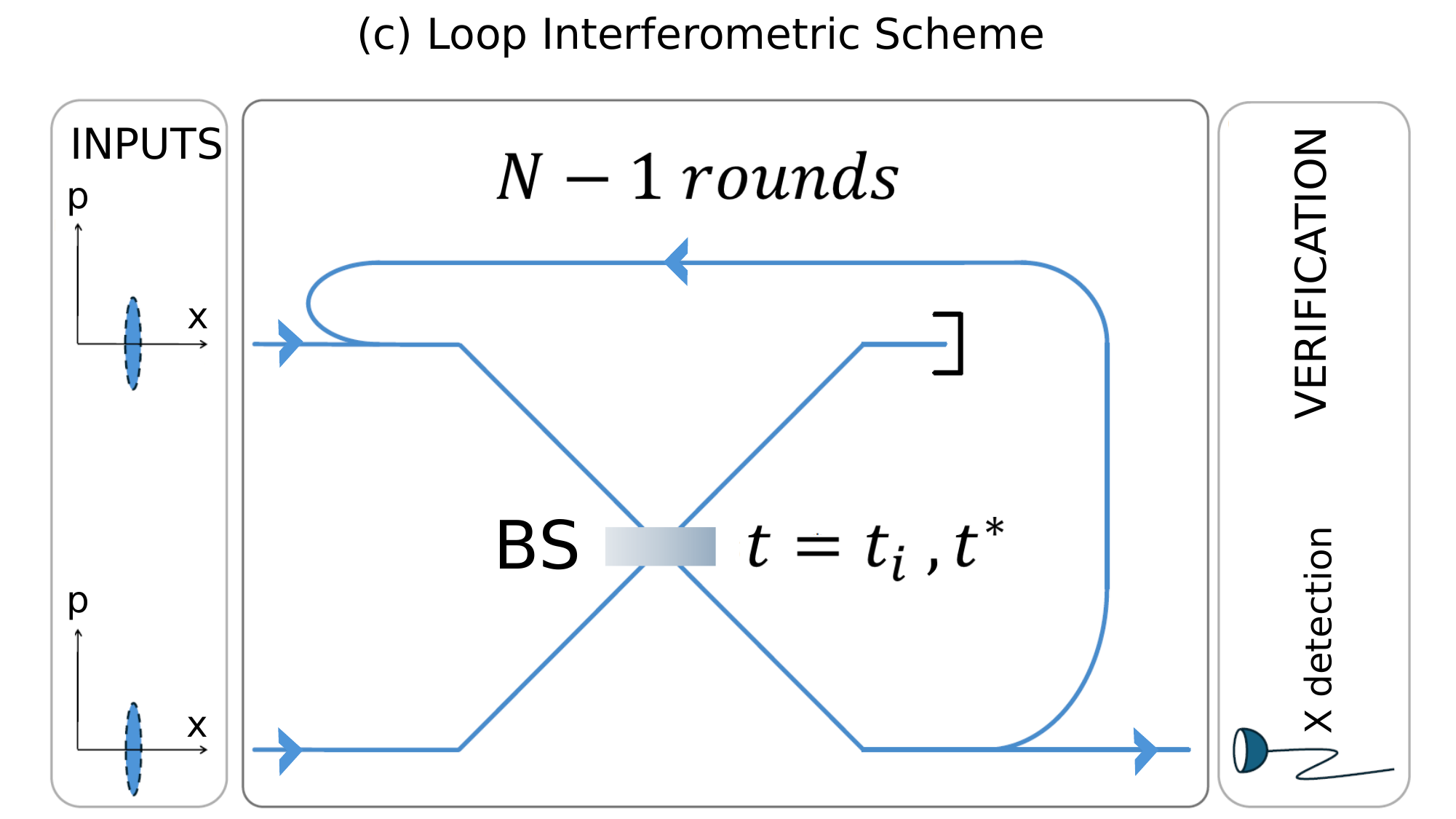} 
	\label{fig:looparchitecture}}
	\end{subfigure}
	\caption{ {Different architectures of interferometric schemes: (a) Pyramidal architecture is an interferometer composed solely of balanced beam splitters. Each pair of input modes interacts, and the constructive output modes of two adjacent beam splitters are utilised as the input modes of the next beam splitter in the interferometer. At the output, this pyramid ensures full constructive interference of mean inputs in the $X$ variable. The output is verified by detecting this variable. (b) Sequential architecture must employ transmittances $t_i = i/(i+1)$ to achieve the same constructive interference in the output as the pyramidal strategy, with identical mean and standard deviation. (c) Dynamical and fixed loop architectures, where the input modes are temporally multiplexed in time-bins, interfere on a single beam splitter with transmittance $t_i$ in such a way that the constructive interference output is re-injected into the first input and interacts with a new squeezed state. The transmittance $t_i$ can vary dynamically as $t_i = i/(i+1)$ to create a fully equivalent output to the pyramidal and sequential architectures. In simpler scenarios, the transmittance is optimised to a single fixed value $t^{*}$ according to the number $N$ of the interfering copies to achieve maximal output mean displacement, or alternatively is simply fixed to $t_i=0.5$. These simpler strategies do achieve maximal mean displacement at the output, however, they can still partially increase the SNR.}}
	\label{fig:differentarchitectures}
\end{figure*}                          

 {\section{Figure-of-merit: SNR}} 

 {Considering the Gaussian approximation of bosonic states, a signal conveying information in a single variable is encoded as a classical coherent displacement. Using interferometric schemes from linear optics, Gaussian noise cannot be reduced, even with linear measurements, feedforward control or postselection. However, the displacements can be accumulated from multiple inputs into a single output by a constructive interfence. This corresponds to a scenario presented in Fig.1a.}
	
 {Gaussian protocols are primarily designed to achieve maximal mean displacement, after which quantum noise is assessed. Therefore, it is essential to compare the mean displacement from a constructive interference with the standard deviation of quantum noise, rather than examining them separately. 
To achieve this, we utilise a widely accepted figure of merit – the signal-to-noise ratio (SNR), which is calculated by dividing the mean displacement by the standard deviation of the noise associated {with the displaced output state}. Increasing SNR is needed to distinguish displacements with greater certainty.  It is an essential operational prerequisite for any multi-copy interference of signals before any application is considered. The SNR for specific variables is {also} more detailed than the overal state fidelity \cite{bowen_tel_2003,fedorov_tel_2021}. When the standard deviation of noise {varies over the copies}, the SNR fluctuates, and these variations are significant for the protocol gain and stability. This becomes more complex if the interference protocol simultaneously utilises multiple copies with varying noise, as in many recent experiments \cite{yoshikawa_invited_2016,yokoyama_ultra-large-scale_2013,podhora_quantum_2022,larsen_deterministic_2021,asavanant_generation_2019,kovalenko_frequency-multiplexed_2021}}.

 {Considering fluctuating SNR as the figure of merit in our analysis, the mean SNR reflects an improvement of the constructive interference in the protocol, generally visulaized in Fig.1(a). The output mean SNR for $N$ inputs must surpass that of a single input to ascertain whether the multicopy inputs are genuinely advantageous. Moreover, the output SNR of nonclassical protocols must exceed that of their classical counterparts. Additionally, scalability requires that the output SNR increases with $N$. To account for fluctuations in SNR caused by instability, the standard deviation of SNR must be lower than its mean with the increasing $N$. Therefore, the gain-to-instability ratio (GIR) can be proposed for SNR following this approach, as introduced in Fig.1(b). Subsequently, multicopy, nonclassical, and scalable criteria must also extend to GIR. Analysing the impact of instability in SNR on the protocol using GIR is essential, the mean SNR cannot specify such instability, as illustrated in Fig. 1(c).} 

 {Before considering any applications involving numerous copies of the Gaussian states, it is essential to evaluate the fundamental scalability of constructive interference under {varying squeezed standard deviation over copies.} The goal is to enhance the SNR, thereby extending previous protocols that involved only two copies \cite{andersen_experimental_2005,lassen_experimental_2010}.  Different architectures of linear interference protocols, as illustrated in Fig. 2, are currently distinctive due to their high feasibility with numerous copies of Gaussian states in laboratories. Therefore, it is possible to demonstrate a scaling of the average SNR under variations in squeezing noise from copy to copy. However, SNR now becomes a random variable whose average does not provide complete information. SNR fluctuations can be diverse, and their behaviour remains entirely unknown and unexplored, both theoretically and experimentally, even in such basic interference protocols. The risk is that these, even minor, fluctuations can expand faster with N than the average SNR. Such protocol functionality can be significantly degraded by slightly {varying squeezed standard deviation} compared to the ideal case. Therefore, a comprehensive analysis must be performed both theoretically and experimentally, which will uncover quantum interference protocols in a new, much more realistic designs. {Only after a thorough understanding and verification of the scalability of interference effects in terms of SNR can the analysis of applications with different figures of merit proceed with comparisons to this essential SNR scalability.}}

 {\section{Interferometric architectures}}

 {To concentrate mean displacements from all inputs into a single output, three configurations illustrated in Fig.\ref{fig:differentarchitectures} are possible in linear interferometry: pyramidal, sequential, and dynamic loop. The pyramidal and sequential architectural configurations consist of $N-1$ beam splitters, whereas the loop architecture employs a dynamically variable beam splitter. If the interferometry is ideal, they are equivalent. However, in realistic, slightly imperfect interferometers, they may differ. In our theoretical analysis, we will consider the case of ideal and stable interferometry, but with realistic and unstable noise in the input states. } 

 {In the pyramid shown in Fig. \ref{fig:pyramidalarchitecture}, one must combine every pair of input modes on a {\em balanced} beam splitter. After this initial layer of interaction, one should collect all the $N/2$ constructive interference output modes from the beam splitter of the first layer and use them as the input modes for the second layer of the balanced beam splitter. This process should be repeated until only one final constructive output mode remains and all the other output modes are traced out.}

 {The sequential architecture in Fig. \ref{fig:sequentialarchitecture} involves the interference of the first two inputs on a balanced beam splitter. Next, the constructive interference output is combined with a new {third} replica of the inputs at an unbalanced beam splitter with a transmittance of $2/3$. This process should be repeated by interfering {the output of with the $(i+1)$th input} on an unbalanced beam splitter with transmittance $t_i = i/(i+1)$. Finally, after interfering all the $N$ inputs on this interferometer, one ideally achieves an output with the same output as the pyramidal architecture (see Fig. \ref{fig:sequentialarchitecture}).} 

 {The dynamic loop architecture illustrated in Fig.\ref{fig:looparchitecture} employs an interference of temporally multiplexed inputs in time-bins at a single beam splitter, while also utilising a varying transmittance $t_i=i/(i+1)$. This design is exceptionally compact but necessitates an additional delay element to ideally achieve the same output as the pyramidal and sequential architectures.}

 {Conversely, simpler architectures can be considered to achieve an output with sufficient SNR and reduced control of the interferometer. They may replace previous architectures, which could be more challenging to implement. For instance, one might envision a loop architecture with time-multiplexed inputs (see again Fig.\ref{fig:looparchitecture}), where the constructive interference output interferes with the subsequent input, but now on a beam splitter with a fixed yet optimised transmittance $t^*$ or by setting the fixed transmittance without optimisation.}

\section{Results: pyramidal, sequential and dynamical loop strategy}

Let us first consider the  multicopy interference  of classical states of bosonic modes.
By interfering two replicas of  {an ideal} coherent state (with position and momentum standard deviations $\Delta x = \Delta p = 1 $ and position mean value $\bar{x}^{(1)} = x_0$) on a balanced beam-splitter interferometer, the state in the constructive interference output mode is a coherent state with displacement  {$\bar{x}^{(2)}=\sqrt{2}x_0$}. For the coherent states with imposed additive classical thermal noise with  $\Delta x_0 = \Delta p_0 = \sqrt{1+\sigma_0^2}$, the displacement increases by the same factor $\sqrt{2}$, but the standard deviation remains $\sqrt{1+\sigma_0^2}$. 

One can extend this architecture to an  {interferometer 
that is feeded with $N$ coherent input states.}
If all $N$ input states are pure coherent states with identical initial displacement $x_i = x_0$ (where the subscript $i$ labels the input mode number), the coherent states in the constructive interference output mode will have a displacement  {
$\bar{x}^{(N)}_{out,cl} 
= \sqrt{N} x_0$.} In the case of perfectly interfering $N$ input coherent states on a linear interferometer constituted of balanced beam-splitters is
\begin{equation}\label{eq:SNRcoherentstable}
	\mbox{SNR}^{(N)}_{out,cl} = \frac{\bar{x}_{out,cl}^{(N)}}{\Delta x_{out,cl}^{(N)}} = \sqrt{N} x_0.
\end{equation}
This  ideal interference of coherent states constitutes our classical benchmark in the scalability of the signal in a passive interferometer. Expectedly, if thermal noise is imposed, the standard deviation remains invariant and only reduces Eq.(\ref{eq:SNRcoherentstable}) by a multiplicative factor $\sqrt{1+\sigma_0^2}$. Moreover, considering noise figure NF=SNR$_{out}^{(N)}$/SNR$_{in}^{(1)}$ between the output SNR and the input SNR after the  interference we gain $\text{NF}_{cl}=\sqrt{N}$ for all classical ideal interference.

However, when the inputs are not stable over the statistical ensemble and slowly drift or diffuse in the signal or noise, the interference may change substantially. To follow physics closely as needed in operational approach, we consider quantum  {optical platform,} broadly known to the bosonic community and with large experimental feasibility of such  {multicopy} protocols  {\cite{marek_general_2018,asavanant_generation_2019,enomoto_programmable_2021,yonezu_time-domain_2023,takeda_-demand_2019}}.  {For nearly ideal classical light from shot-noise limit laser only the signal amplitude $x_0$ can vary at longer time scale following a Gaussian distribution due to thermal pump changes {\cite{scully_quantum_1997}}. The stabilized standard deviation $\Delta x=1$ is practically unaffected.} Such a stable shot-noise limit is essential for calibrating the scale of position and momentum  quadrature measurements \cite{bachor_guide_2009}.

By taking the instability in amplitude into account, we assume the input signal is in the form $x_i = x_0 + \delta x_i$, where $\delta x_i$ is a random variable following a centered normal distribution {$\mathcal{N}(0,\sigma_{x_0}^2)$ where $\sigma_{x_0}^2$} is the variance in the signal distribution. The SNR averaged on the displacement fluctuations reduces to the classical SNR in Eq.(\ref{eq:SNRcoherentstable}) as:
$
\mean{SNR^{(N)}_{out,cl}}_{\delta x_i} = \sqrt{N} \left( x_0 + \mean{\delta x} \right) = \sqrt{N} x_0,
$
where $\delta x = (\delta x_1,..., \delta x_i, ..., \delta x_N)^T$ is the signal fluctuation vector.
Hence, the fluctuations in signal are statistically irrelevant  {for mean SNR} in the regime of large $N$.
If the laser is not shot-noise-limited  {but $\Delta x_0$ is stabilized,} it follows the same pattern and finally, such classical multimode interference instability in the signal has  {always} $\text{NF}=\sqrt{N}$.

 {Nonclassical} squeezed states can be used to carry information encoded in the displacement with smaller noise. The replacement of coherent states by displaced squeezed states with a stable noise as input of our linear interferometer will not affect the mean displacement in the protocol which follows the one of the classical case. However, the standard deviation in the $x$-quadrature of the output mode will be reduced by a factor of  {$e^{-r}$,} where $r$ is squeezing parameter proportional to pump amplitude.  {Compared to the coherent state variance, the squeezing leads to } 
\begin{equation}\label{eq:SNRsqueezedstable}
		\mbox{SNR}^{(N)}_{out,sq} = \frac{\bar{x}_{out,sq}^{(N)}}{\Delta x_{out,sq}^{(N)}} = \sqrt{N} e^{r} x_0.
\end{equation}
Hence, the effect of using stable squeezed states is to increase the classical SNR from Eq.(\ref{eq:SNRcoherentstable}) by a multiplicative factor $e^r$ while keeping the scaling of $\sqrt{N}$ unaffected reflected by an ideal noise figure $\text{NF}=\sqrt{N}$,  {still same} as for the classical case.
As the squeezing of the input mode states is stable ($r_i = r$ for all $i$'s), then Eq.(\ref{eq:SNRsqueezedstable}) is the upper bound on the SNR that can be reached. 
	
However, when the shot-noise-limited laser light with  {an instability in amplitude} discussed above is used as the pump in optical parametric oscillator (OPO) to generate squeezing \cite{scully_quantum_1997},  {the squeezing level can vary}. For squeezed states  {with $\Delta x = e^{-r}$,} where $r$ is the squeezing parameter  linearly depending on the  {uncertain} amplitude of the pump \cite{scully_quantum_1997}, SNR and NF begin to change substantially due to  {their} averages of the exponent of  {varying} $r$. These results are in our example obtained by modelling the effect of laser amplitude instability on the squeezing properties of the output squeezed states in a  {$\chi^{(2)}$} squeezing process typically used in optical parametric amplifiers. Note that the results might be different if the squeezing process is made by a  {$\chi^{(3)}$} process like four-wave-mixing  \cite{glorieux_hot_2023,vaidya_broadband_2020,zhang_squeezed_2021}, as the squeezing is proportional  {to the} amplitudes of two pumps. {To demonstrate our methodology, we model that the only source of instability is the squeezing parameter. Phase instability affects both classical coherent and non-classical states, such as squeezed-displaced states, and is significant for certain experiments. Nonetheless, these instabilities can be mitigated through phase-locking with a classical light beam. However, such techniques do not address the instability of the output coherent amplitude and squeezing from a parametric oscillator or amplifier, which typically differ from pump fluctuations. Therefore, we initially concentrate on these aspects to demonstrate the approach in this manuscript. Future experimental verification might incorporate phase fluctuations; however, including them would complicate the presentation of the method. In our model of interference of squeezed states, the effect of a pure-loss on the input modes is to reduce the SNR but it increase the GIR as it stabilizes the squeezing fluctuations (see Fig. \ref{fig:GIRsqueezingLosses}). We refer the interested reader to appendix \ref{app:losses} for details on the effect of losses on the SNR and GIR.}
	
Therefore, in a more realistic description of squeezing from the OPO, one need to take into account the amplitude instability of the pump laser for the squeezing generation that translates in \textit{independent} changes $\delta r_i$ of the squeezing parameter $r_i$ of each input mode around a stable value $r_0$. Hence, the $N$ replicas of input pure squeezed states are considered  to have identical signal values $x_i = x_0$ but independently varying squeezing parameter value $r_i = r_0 + \delta r_i$ where, similarly to the signal fluctuations, the $\delta r_i$'s are  assumed to be distributed from a Gaussian distribution $\mathcal{N}(0, \sigma_r^2)$. {This assumption is relevant in experimental situations where $\sigma_r$ is independent of $N$, therefore, we have no prior limits to scale it up until the replicas interfere.} The input signal-to-noise ratio, as our figure of merit, is defined as 
$\mbox{SNR}_{in,i}=x_0\exp{r_i}$
where $r_i$ is the squeezing parameter varying for each copy. Already the single-copy average $\mean{\mbox{SNR}_{in}^{(1)}} = x_0\exp{(r_0)}\exp{(\sigma_r^2/2)}$ exponentially increases with the variance $\sigma_r^2$ of that Gaussian fluctuations.
 {However, the standard deviation $\Delta\mbox{SNR}_{in}^{(1)}=x_0\exp{(r_0)}\exp{(\sigma_r^2/2)}\sqrt{\exp{(\sigma_r^2)}-1}$ and therefore the uncertianty in SNR grows faster than its mean value and there is no real positive gain. It suggests that comparison of mean SNR and its standard deviation is important to understand the results. To make comparison quantitiative, a stability ratio $\mean{\mbox{SNR}_{in}^{(1)}}/\Delta\mbox{SNR}_{in}^{(1)}=1/\sqrt{\exp{(\sigma_r^2)}-1}$ describes purely the effect of the {varying squeezing standard deviation} irrespectively to $x_0$ and $r_0$.  
Note for later, for the small $\sigma_r$, we take the series expansion to the first order in $\sigma^2_r/2$, and obtain  $\mean{\mbox{SNR}_{in}^{(1)}} \approx x_0(\exp{r_0})(1+\sigma^2_r/2)$ and $\mean{\mbox{SNR}_{in}^{(1)}}/\Delta\mbox{SNR}_{in}^{(1)}\approx 1/\sigma_r - \sigma_r/4$.} 

In this scenario  {with varying squeezing over the copies}, the SNR 
at the constructive output of the ideal beam-splitter network  {is  
\begin{equation}\label{eq:SNRsqueezingfluctuation}
	\mbox{SNR}^{(N)}_{out,sq} = \frac{\sqrt{N} x_0 }{\sqrt{\frac{1}{N} \sum_i^N e^{-2  r_i}}} = \frac{N e^{r_0} x_0 }{\sqrt{\sum_i^N e^{-2  \delta r_i}}},
\end{equation}}
by assuming  {the accumulated mean displacement in the nominator without any change as it is stable or with only additive Gaussian noise as described above.} Generally, the  {uncertainty in Eq.(\ref{eq:SNRsqueezingfluctuation}) over $\delta r_i$} can be analysed only numerically.
However, by assuming that the fluctuations in the squeezing parameters $\delta r_i$ are small (meaning that {$\sigma^2_r \ll 1$}), we can approximate Eq.(\ref{eq:SNRsqueezingfluctuation}) in series up to the second order in $\delta r_i$'s and take the average. Note that the $\delta r_i$'s are considered to be independent and identically distributed from the normal centered distribution. Hence, they are assumed to have zero-mean $\mean{\delta r} = 0$ and to be uncorrelated. The approximated  {mean SNR from Eq.(\ref{eq:SNRsqueezingfluctuation})} is
	\begin{equation}\label{eq:SNRsqueezingapprox}
		\mean{\mbox{SNR}_{out,sq}^{(N)}}
		\approx  \sqrt{N} e^{r_0} x_0 \left(1 + \left(\frac{3}{2 N} - 1 \right) \sigma_r^2 \right),
	\end{equation}
	which in the asymptotic limit of large $N$ is converging to:
	\begin{equation}\label{eq:SNRsqueezingAsymptotic}
		\mean{\mbox{SNR}_{out,sq}^{(N)}} \rightarrow \sqrt{N}  e^{r_0} x_0 \left(1 -  \sigma_r^2 \right)
	\end{equation}
 {that can be compared with input $\mean{\mbox{SNR}_{in}^{(1)}} \approx x_0(\exp{r_0})(1+\sigma^2_r/2)$.} 	
Remarkably,  {by the constructive interference,} the small random differences in squeezing of the input modes only slightly and linearly modifies the ideal mean output SNR by a multiplicative prefactor $\left(1 -  \sigma_r^2 \right)$. This is  {also visible in only small decrease of} $\mbox{NF} = \sqrt{N} \left(1 -  \sigma_r^2 \right)$. Hence the scaling factor $\sqrt{N}$  {for the mean SNR} is preserved for $\sigma_r$ significantly smaller than $1$.

 {However, similarly as for the input SNR, we must compare the average  $\mean{\mbox{SNR}_{out}^{(N)}}$ from Eq.(\ref{eq:SNRsqueezingfluctuation}) with the uncertainty in such SNR by a standard deviation  $\Delta \mbox{SNR}_{out}^{(N)} = \sqrt{\mean{(\mbox{SNR}_{out}^{(N)})^2} - \mean{\mbox{SNR}_{out}^{(N)}}^2}$ to understand our results and their scaling. Then, their ratio will determine the role of the squeezing instability. The larger is such ratio, smaller is the effect of the squeezing instability on the SNR determined by Eq.(\ref{eq:SNRsqueezingfluctuation}). We utilise the standard deviation in the denominator of GIR to obtain a conservative characteristic of the {varying} resources.} 

Motivated by this, we propose a general outline of this approach. Having a figure-of-merit $F$ for the protocol, we consider the \textit{average gain} $\langle F_{out}^{(N)}\rangle$ using $N$ copies of resources  {that are not stable}. We define a \textit{gain-to-instability ratio} (GIR) by normalizing the gain by a standard deviation $\langle \Delta F_{out}^{(N)}\rangle$. {The closest notion to the GIR in classical signal processing is, to our knowledge, the Amount of Fading (AoF) of the instantaneous SNR encountered for example in the field of wireless communication \cite{cheng_bivariate_2020}. In the fields of quantum information and communication, fading has been described in \cite{dong_experimental_2008,dong_continuous_2010,
ruppert_fading_2019,usenko_entanglement_2012} where unstable transmittivity has been used, but GIR was not defined, considered nor used.}
In the unrealistic case of ideally stable identical copies, the GIR diverges, similarly to the SNR in the unrealistic case of no fluctuations. Analogically, as for the signal-to-noise ratio in classical signal analysis, $\mbox{GIR}>1$ is at least needed, and a faster increase with $N$ makes scaling of averaged gain more reliable. A saturation of GIR limits the scalability of reliable protocols under the varying resources, although the average gain scales further with $N$. Therefore, although different protocols and their architectures might have a nearly comparable scaling of  {the gain in $\langle F_{out}^{(N)}\rangle$ over $N$ copies,} they may differ substantially in the gain reliability quantified by the GIR.  {Also the oppposite situation can happen, the architectures might have different gain in $\langle F_{out}^{(N)}\rangle$ over $N$ copies, but GIR might suggest similar gain reliability.} 

Let us continue our guiding example with  {the varying squeezing and figure-of-merit being SNR.}
Although such  {GIR} is essential, its full evaluation is also possible only numerically, even in our simple case, making the outcome  {not simply predictable and} nontrivial.  {For the pyramidal, sequential and dynamical loop architectures, the GIR can be approximately evaluated in the second-order expansion to predict} 
	\begin{equation}\label{eq:SNRSNRsqueezingPyr}
		\mathrm{GIR_{SNR}} = \frac{\mean{\mbox{SNR}_{out,sq}^{(N)}}}{\Delta \mbox{SNR}_{out,sq}^{(N)}}
		\approx \sqrt{N} \left( \frac{1}{\sigma_r} + \left(\frac{3}{2 N} - 1 \right) \sigma_r  \right),
	\end{equation}
which complements the approximative formula for mean $\mbox{SNR}$ given by Eq. (\ref{eq:SNRsqueezingapprox}). Differently, the standard deviation $\sigma_r$, not the variance $\sigma_r^2$, determines Eq. (\ref{eq:SNRSNRsqueezingPyr}). In the limit of very large number  {$N$ of input modes, (\ref{eq:SNRSNRsqueezingPyr}) converges to} 
	\begin{equation}\label{eq:SNRSNRsqueezingPyrAsympt}
		\mathrm{GIR_{SNR}} =\frac{\mean{\mbox{SNR}_{out,sq}^{(N)}}}{\Delta \mbox{SNR}_{out,sq}^{(N)}}  \rightarrow \sqrt{N} \left( \frac{1}{\sigma_r} -  \sigma_r \right).
	\end{equation}
First, let us appreciate that  {the scaling in  Eq.(\ref{eq:SNRSNRsqueezingPyrAsympt}) depends of the number of input states $N$ exactly like Eq.(\ref{eq:SNRsqueezingAsymptotic}) for the mean SNR. On the other hand, 
GIR in Eq.(\ref{eq:SNRSNRsqueezingPyrAsympt}) does not depend on the input mean displacement $x_0$ and squeezing $r_0$ similarly as the noise figure $\mbox{NF}=\sqrt{N}(1-\sigma_r^2)$.}
Second, crucially, gain-to-instability ratio Eq.(\ref{eq:SNRSNRsqueezingPyrAsympt}) decreases inversely to $\sigma_r$, that makes reduction much faster than the average SNR Eq.(\ref{eq:SNRsqueezingapprox}). Therefore, increasing the number $N$ of input copies allows to compensate such decreases much slowly  {than for the mean in Eq.(\ref{eq:SNRsqueezingAsymptotic}) and corresponding NF$=\sqrt{N}(1-\sigma^2_r)$ and irrespectively to $x_0$ and $r_0$. Considering $\sigma_r=0.1$ and $\sigma_r=0.2$ and limit of large $N$, NF per a single copy reduces negligibly only from $0.99$ to $0.96$, however, GIR drops more from $9.9$ to $4.8$. The later case with $\sigma_r=0.2$ requires four times more copies to reach the same GIR as the former one with $\sigma_r=0.1$. GIR is much more demanding to compensate by more copies for these architectures than NF. Therefore, GIR must be evaluated to complement analysis of the gains in the multicopy squeezing protocols.} 

\section{Discussion: fixed loop strategies, numerical analysis and measurement-induced squeezing}

\begin{figure*}[ht!]
\centering
\begin{subfigure}{
 \includegraphics[width=0.48\textwidth]{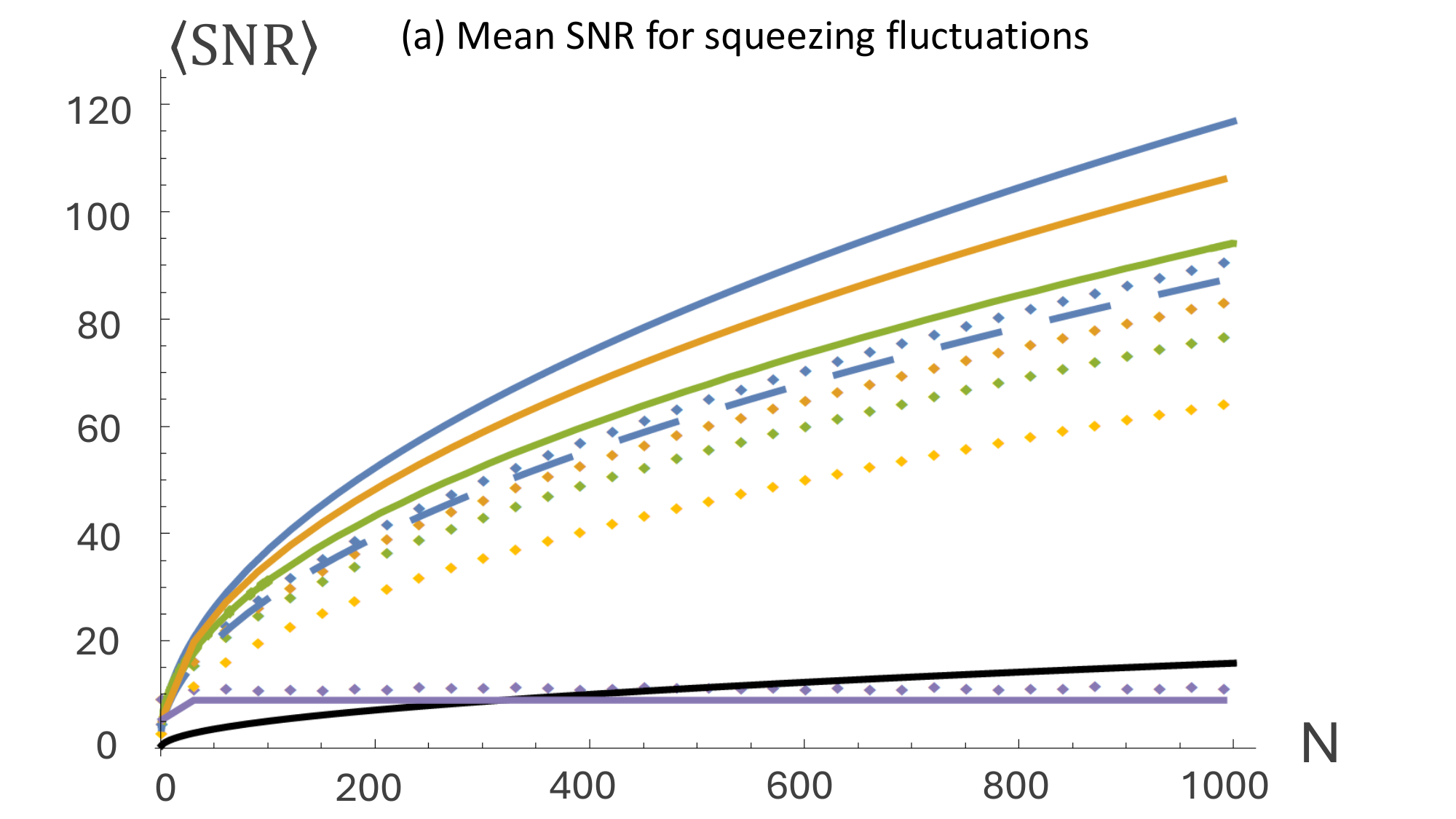}\label{fig:SNRsqueezing}}
 \end{subfigure}
 \begin{subfigure}{
 \includegraphics[width=0.48\textwidth]{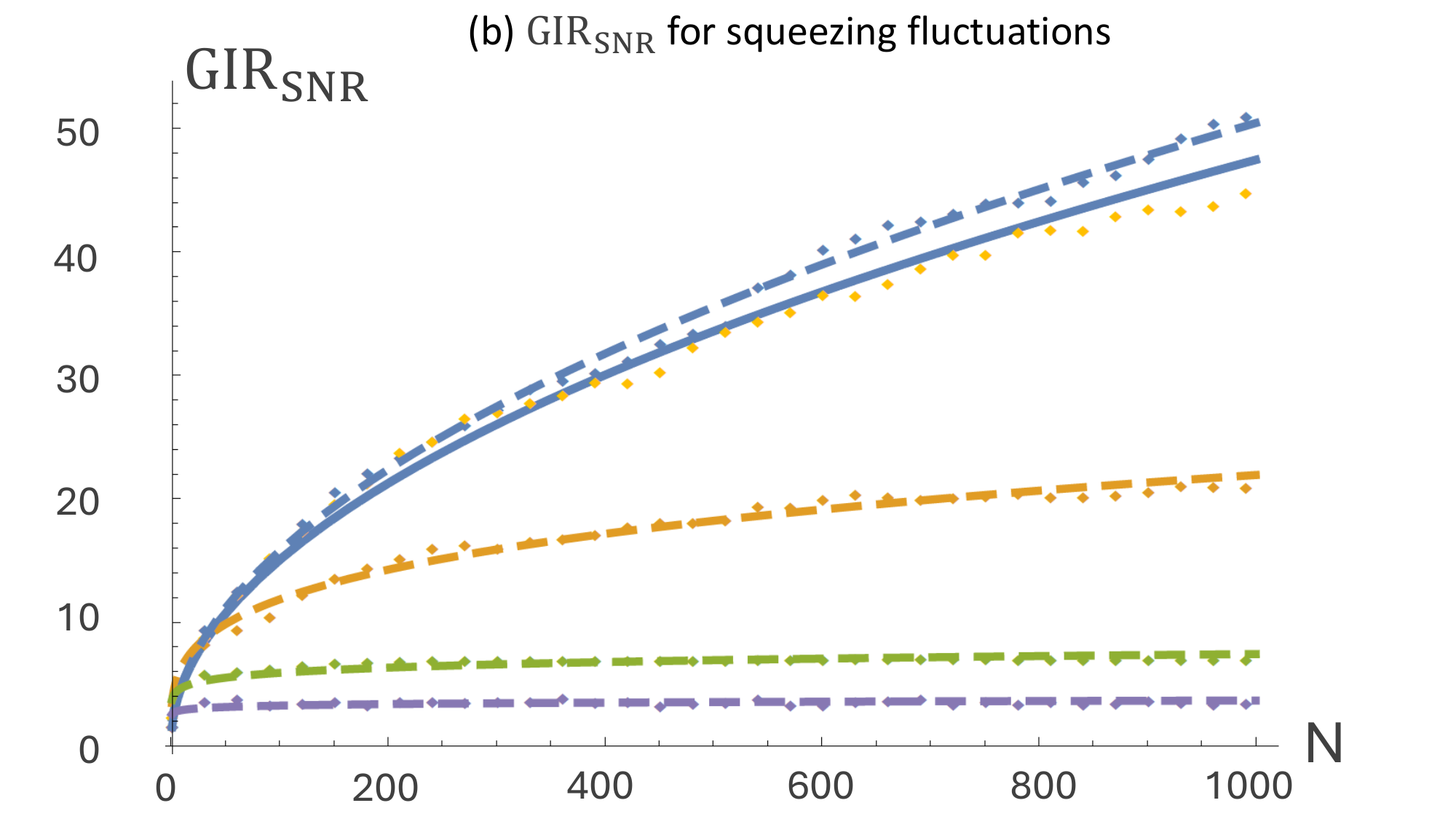}\label{fig:SNRSNRsqueezing}}
\end{subfigure}
\caption{Constructive interference of $N$ equally displaced squeezed states with {varying standard deviation of noise} for the different architectures. 
The parameter values are the same for all architectures and are set to $x_0 = 0.5$, $r_0 = 2$, $\sigma_r = 0.5$.
(a) The mean signal-to-noise ratio $\langle \mbox{SNR}\rangle$: continuous lines represent the ideally stable squeezing for each architecture. The dotted lines account for the squeezing instability. The blue line corresponds to the pyramidal, sequential, and dynamical loop architectures as defined by Eq.(\ref{eq:SNRsqueezedstable}). For the fixed loop architecture outlined in Eq.(\ref{eq:SNRSqueezingLoop}), the orange line denotes a transmittance $t^*$, the green line represents a fixed transmittance $t_{i}=(N-1)/N$, while the purple line indicates the fixed $t_i=0.5$, independent of $N$. The blue dotted line relates to Eq.(\ref{eq:SNRsqueezingfluctuation}), and for comparison, the blue dashed line is provided by the approximate Eq.(\ref{eq:SNRsqueezingapprox}) (except for very small $N$, which is not shown on the graph). The dotted orange, green, and purple lines adhere to Eq.(\ref{eq:SNRSqueezingLoop}) for $t=t^*$, $t_i=1-1/N$, and $t_{i}=0.5$, respectively. For comparison, the yellow points indicate the average SNR for measurement-induced displacements from varying two-mode squeezed states {\cite{fiurasek_conditional_2001,schnabel_squeezed_2017}}. The black curve represents the classical limit as given by Eq. (\ref{eq:SNRcoherentstable}), where the input states are coherent states.
(b) The gain-to-instability ratio $\mbox{GIR}_{\mbox{SNR}}=\langle \mbox{SNR} \rangle/\Delta \mbox{SNR}$: The blue points correspond to the GIR from numerical simulations, and the blue dashed line represents their exponential fit by $1.5 \times N^{0.51}$. For comparison, the blue continuous line is the approximate Eq.(\ref{eq:SNRSNRsqueezingPyr}); its asymptotic behaviour, as described in Eq.(\ref{eq:SNRSNRsqueezingPyrAsympt}), is indistinguishable from the blue continuous line. {The yellow points correspond to the GIR for measurement-induced displacements from {varying} two-mode squeezed states.} The orange points represent the GIR numerically simulated for the loop architecture with optimised transmittances $t = t^*(N)$. The dashed orange line is their exponential fit $3.5 \times N^{0.27}$. Finally, the green points curve represents the data for the loop architecture with transmittance fixed at $(N-1)/N$, which saturates quickly. It is similar to the purple curve for the loop architecture with $t=0.5$, which performs poorly. Note that all the parameters and exponential fitting values are in Tables \ref{tab:SNRfit} and \ref{tab:GIRfit}; see Appendix \ref{appendixD:fitparametervalues}.}
\label{fig:3}	
\end{figure*}
	
The pyramidal, loop and dynamical loop architectures lead to an SNR scaling law proportional to $\sqrt{N}$ with a  {$\mbox{GIR} =\mean{\mbox{SNR}}/\Delta \mbox{SNR}$} that also scale as $\sqrt{N}$,  as visible in Fig.\ref{fig:3}. However, the experimental setup to realize  {pyramidal and sequential architecture use of $N-1$ beam splitters and the dynamical loop architecture needs to dynamically vary the single beam splitter. Therefore, it is worth to investigate} a more experiment-friendly loop architecture consisting of only single beam-splitter with some fixed transmittance $t$ that can be optimized respective to the number $N$ of copies  {to reach maximal mean displacement at the output. For the $N$ input copies, we approach}
	\begin{equation}\label{eq:SNRSqueezingLoop}
		\mbox{SNR}^{(N)}_{out,sq} = \frac{x_0\sum_{i=1}^N \tau_i^{\frac{1}{2}}}{e^{-r_0}(\sum_{i=1}^N \tau_{i} e^{-2 \delta r_i} )^{\frac{1}{2}}},
	\end{equation}
	where 
	\begin{equation}\label{def:tau}
		\tau_i =
		\left\{
		\begin{array}{ll}
			t^{N-1}  & \mbox{if } i = 1, \\
			(1-t) t^{N-i} & \mbox{if } i > 1
		\end{array}
		\right.
	\end{equation}
 {are accumulated transmittances at the output. Note that in the ideal case without fluctuations, Eq.(\ref{eq:SNRSqueezingLoop}) denominator is identically equal to one and, therefore, only mean displacement in numerator is optimized.}

The values of the transmittance $t=t^{*}$ become very close to one for large $N$  {and can be numerically obtained from the maximum of approximative mean displacement $x_0\sqrt{1-t^{*}}(\sqrt{t^{*}}-t^{*N/2})/(\sqrt{t^{*}}-t^{*})$ determining nominator of (\ref{eq:SNRSqueezingLoop}).
Approximative $t^{*}\approx 1-2.4/N$ is able to reach the mean displacement $0.9x_0\sqrt{N}$, i.e. 90$\%$ of the displacement $x_0\sqrt{N}$ of the pyramidal, sequential and dynamical loop architectures. For small $\sigma_r^2\ll 1$, the approximated $\mean{\mbox{SNR}_{out,sq}^{(N)}}$ approaches $0.9 \sqrt{N}  e^{r_0} x_0 \left(1 -  \sigma_r^2 \right)$ lowered by the smaller mean displacement. Although only a single beam splitter is sufficient, a precise setting of the transmittance close to unity is still needed for large number of input modes $N$.}

 {To simplify strategies further, one can fix transmittance of the beam splitter without such optimization. We have taken two possible values to be a fixed balanced transmittance $t_1 = 1-1/N$, and $t_2 = 1/2$ for a comparison, both corresponding to the last and first beamsplitter for the sequential and dynamical loop strategies. In the former case of $t_1 = 1-1/N$, we can approximately reach mean displacement $0.79x_0\sqrt{N}$ still 80$\%$ close to pyramidal, sequential and dynamical loop method and also $\mean{\mbox{SNR}_{out,sq}^{(N)}}$ changes satisfactorily by the same factor. In the later case, it is quickly saturated over $N$ to maximally $x_0(\sqrt{2}+1)$, being negligible to all previous cases and inefficient. }
		
In order to test these analytical results  {and, mainly, to predict GIR for different simpler strategies,} we simulate the resulting first and second moment at the constructing interference output of our linear interferometer and study their scaling up properties. These numerical results are performed by sampling the squeezing parameter $r$ from a normal distribution with some fixed standard deviation $\sigma_r$ and mean value $r_0$. Hence, the squeezing parameter for each input pure squeezed state is $r_i = r_0 + \delta r_i$ where the $i$ labels the input mode number and $\delta r_i \sim \mathcal{N}(0, \sigma_r)$. By using Eq. (\ref{eq:SNRsqueezingfluctuation}) and Eq.(\ref{eq:SNRSqueezingLoop}), one can simulate the  {varying} output SNR  {in order to estimate the average $\mean{\mbox{SNR}}$,} for the different architectures.
		
The simulation is carried out for each number $N$ of input state from $N = 2$ to $N = 1000$ and repeated $100$ times to access the $\mean{\mbox{SNR}}$ for each $N$. In  {Fig.\ref{fig:SNRsqueezing},} each diamond point is generated numerically as described.
One can see that the result displayed in  {Fig.\ref{fig:SNRsqueezing}} show that the behaviour of the SNR for the  {pyramidal, sequential and dynamical loop architectures (in blue) and fixed loop architecture with optimized transmittivity  {$t^*$} (in orange) and  {$t_1=1-1/N$} (in green)}  are very similar. Indeed, the scaling law of $\sqrt{N}$ is conserved for  {all} architecture and only the pre-factors are affected by the choice of architecture.  {However,} the difference between the ideal case without any squeezing instability (solid blue line) and the numerical evaluations  of Eq.(\ref{eq:SNRsqueezingfluctuation}) including moderate squeezing fluctuations $\sigma_r=0.5$ (dotted blue line) is already noticeable, the scaling with $\sqrt{N}$ remains  {and only prefactors change when $t=t^*$ and $t=t_1$.} 
 {On the other hand, for the loop with fixed balanced transmittance $t_2=0.5$ (in purple), the SNR is low, saturated due to previosly analysed mean displacement and, therefore, breaks the classical limit at same $N$ (solid black line). The numerator simply has converging series over $N$ for any constant $t$ explaining the unscalability of such fixed transmittivity architecture and need for $t$ depended on $N$.}  

We  {also} point out that the numerical data (dotted blue in  {Fig.\ref{fig:SNRsqueezing}}) from Eq.(\ref{eq:SNRsqueezingfluctuation}) and the approximated SNR (dashed blue line in Fig.\ref{fig:SNRsqueezing}) from Eq. (\ref{eq:SNRsqueezingapprox}) are in good agreement.  {However, such agreement can strongly depend on larger values of the $\sigma_r$ describing squeezing instability.}  {Still, the gain in SNR from even unstable squeezed fluctuations in the pyramical, sequential and dynamical loop architecture is not critically reduced and remains significantly higher than the classical limit (solid black line).} Indeed, even under its instability, the squeezing substantially contribute to the enhancement of the SNR compared to the classical case. 

\begin{figure*}[!ht]
	\centering
	\begin{subfigure}{
	\includegraphics[width=0.48\textwidth]{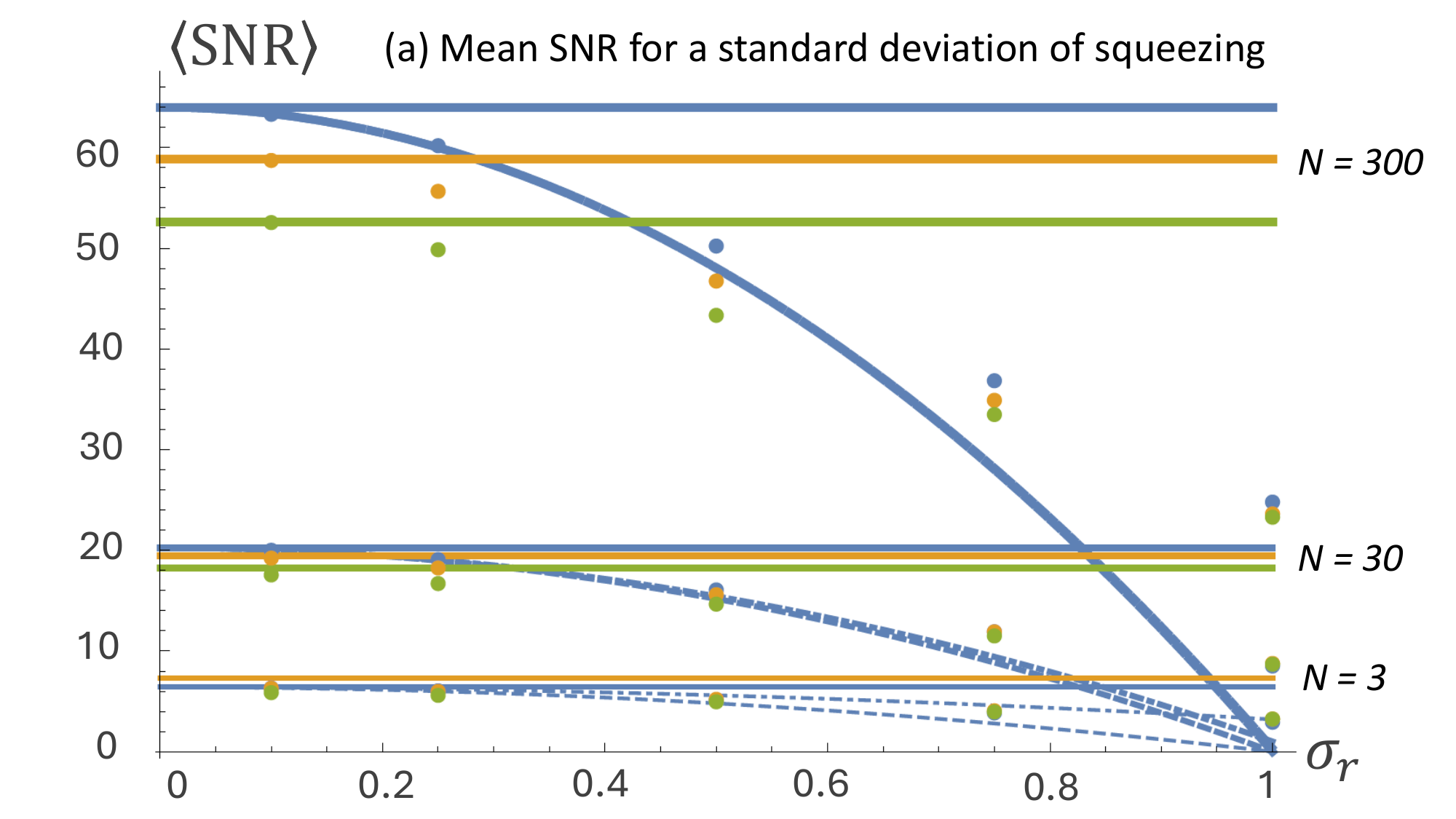} \label{fig:SNRvssigma}}
	\end{subfigure}
	\begin{subfigure}{
	\includegraphics[width=0.48\textwidth]{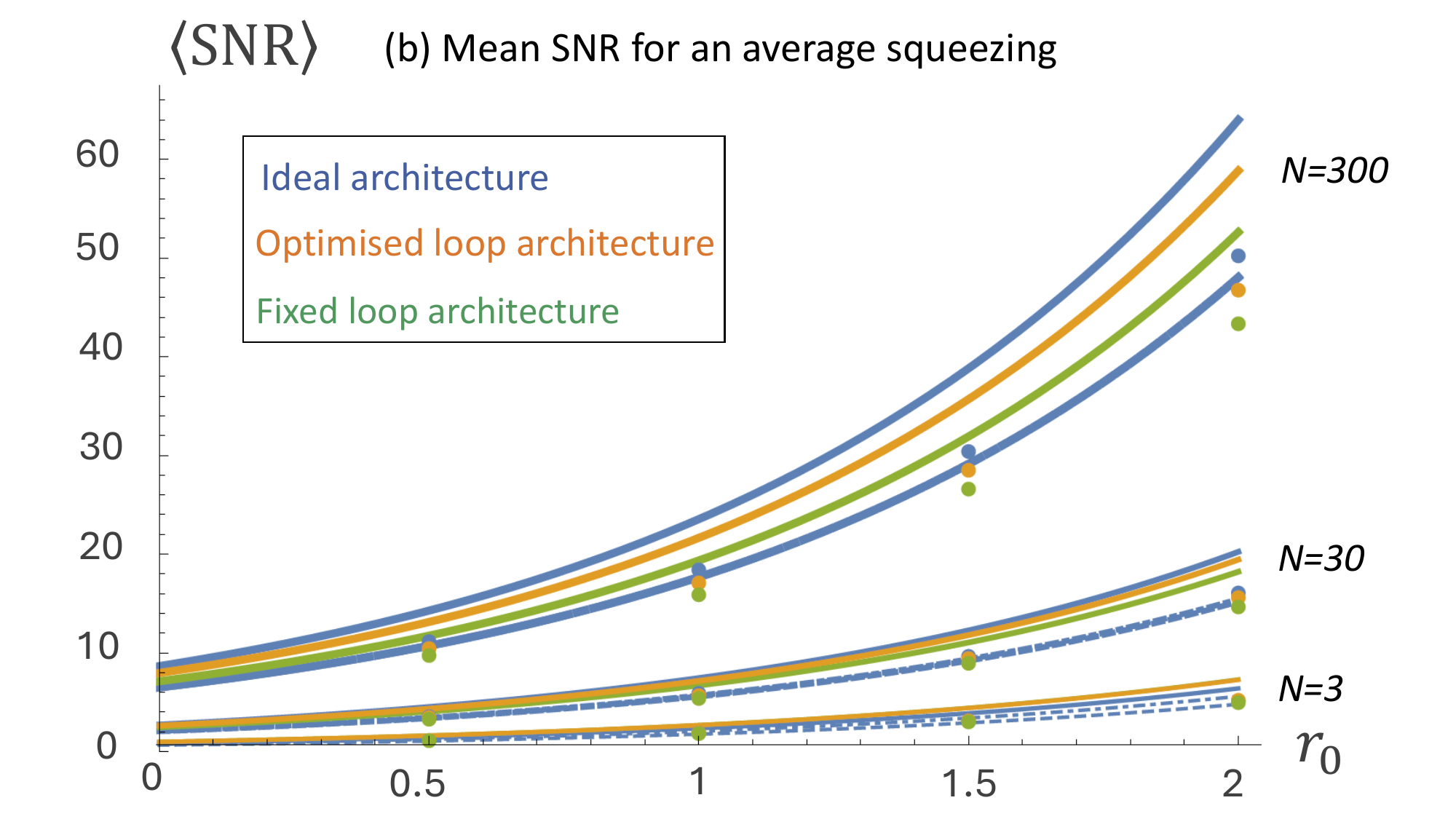} \label{fig:SNRvssqueezing}}
	\end{subfigure}
	\begin{subfigure}{
	\includegraphics[width=0.48\textwidth]{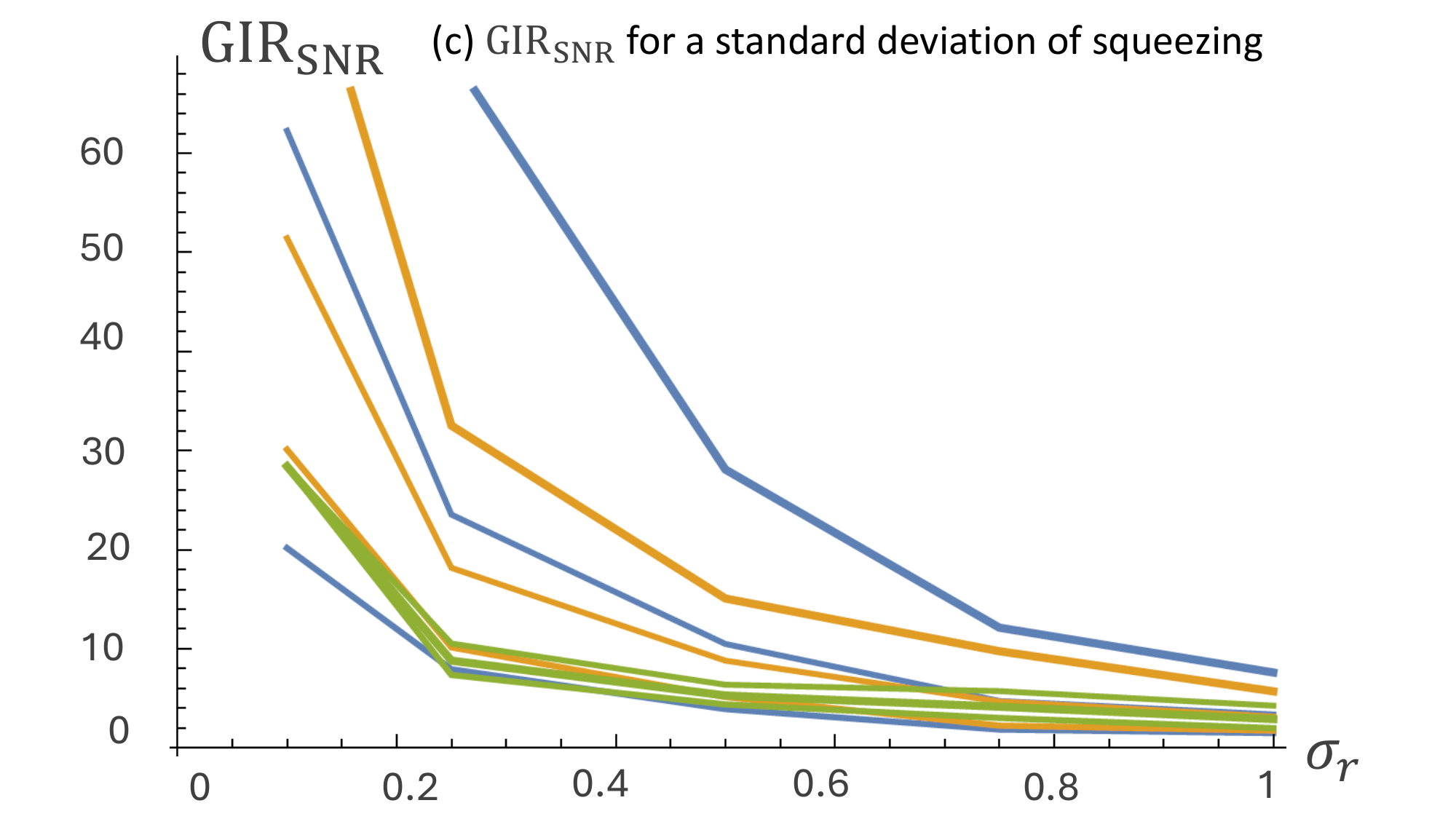} \label{fig:SNRSNRvssigma}}
	\end{subfigure}
	\begin{subfigure}{
	\includegraphics[width=0.48\textwidth]{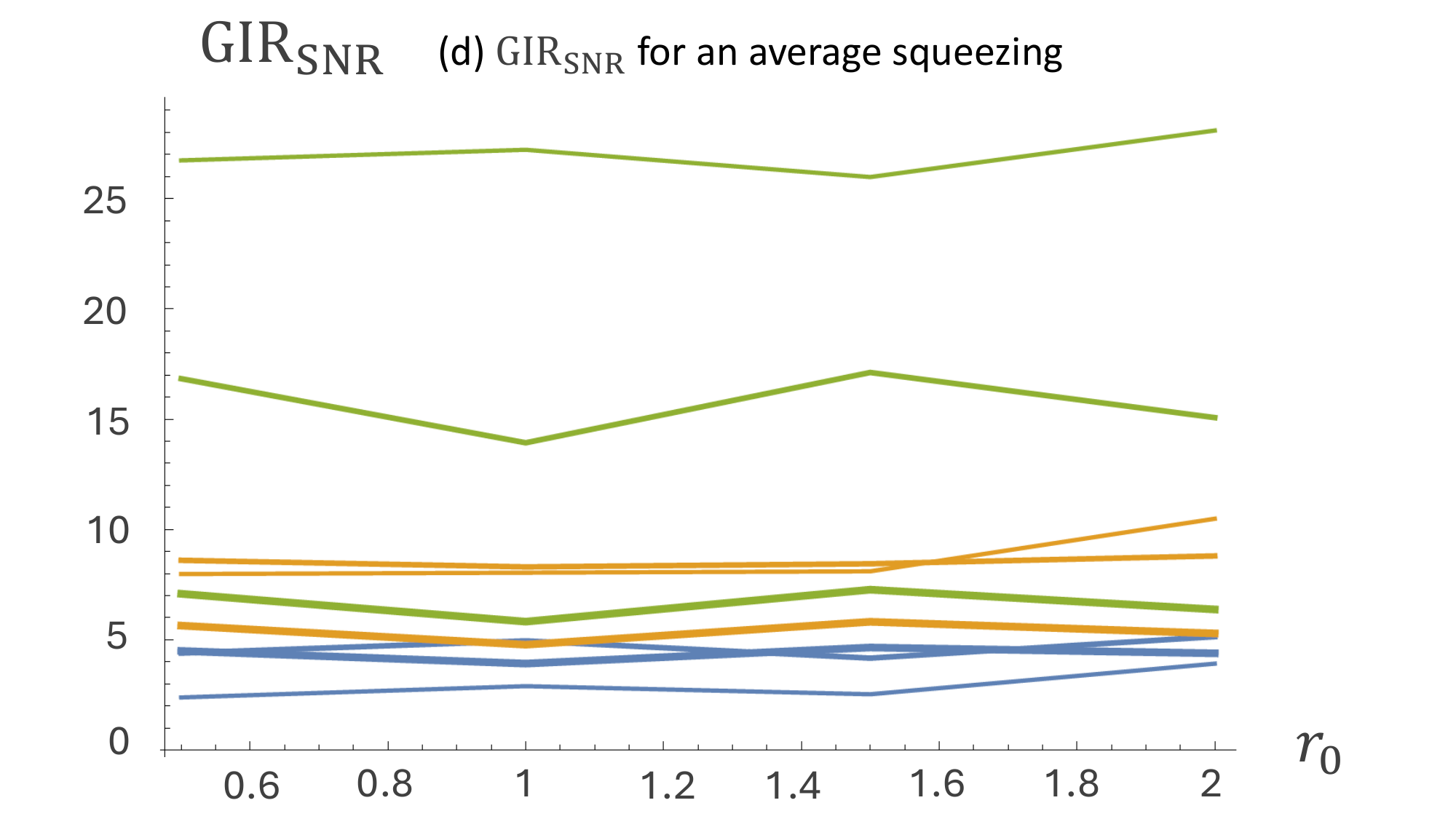} \label{fig:SNRSNRvssqueezing}}
	\end{subfigure}
	\caption{ {Constructive interference in the different architectures at Fig.2 as functions of (a,c) squeezing instability $\sigma_r$ and (b,d) average squeezing $r_0$. Parameters are $x_0 = 0.5$, $r_0 = 2$, $\sigma_r = 0.5$ as in Fig.\ref{fig:3}, unless free parameter. Note that the colors and thickness codes used are the same in all the panels of Fig.\ref{fig:4}. In Fig.\ref{fig:SNRvssigma}, the horizontal continuous blue line is for ideal behaviour in the pyramidal, sequential and dynamical loop architectures, orange lines corresponds to the loop architecture the optimized transmittance $t^*$ and green lines are for fixed transmittance equal $(N-1)/N$. Trios of lines are for $N= 3$ (thinner lines), $N= 30$ and  $N= 300$ (boldest lines) input copies (from bottom to top lines). The points of corresponding colours are the results of numerical simulations for different strategies. The blue decreasing lines are for approximated (Eq. (\ref{eq:SNRsqueezingapprox}); dotdashed line) and asymptotic (Eq. (\ref{eq:SNRsqueezingAsymptotic}); dotted line) formulas. Note that when $N$ increase the approximated and asymptotic formula's for SNR become arbitrarily close and appears like a continuous line. This means that the asymptotic behaviour of the SNR is reached already for a number of modes $N$ of order $10$. In Fig.\ref{fig:SNRvssqueezing}, the colors and thickness codes are used for different architectures (colors) and number of input copies (thickness) for $N=3$ (bottom thin lines), $N=30$ (middle medium lines) and $N=300$ (top bold lines). The dots shows the numerical simulations. For the blue colors,  approximated (dotdashed) and asymptotic (dotted) formulas are displayed. Note that when $N$ increase the approximated and asymptotic formula's for SNR become arbitrarily close and appears like a continuous line for $N=300$.
	For Fig.\ref{fig:SNRSNRvssigma}, the GIR is showed for different architectures (colors) and different number of input modes (thickness) as a function of the standard deviation $\sigma_r$. Here, only the numerical data is displayed. The dots are joined by a continuous line in order to emphasize the decreasing of the GIR with the standard deviation $\sigma_r$. Finally, in Fig.\ref{fig:SNRSNRvssqueezing}, the GIR is showed for different architectures (colors) and different number of input modes (thickness) as a function of the mean squeezing $r_0$. Here again, only the numerical is displayed and the dots are joined by straight lines in order to emphasize their behaviour. Here, one can see that, for all architecture (colors) and number of input modes $N$ (thickness), the GIR is not very sensitive to the mean squeezing $r_0$.}}
\label{fig:4}
\end{figure*}

In Fig.\ref{fig:SNRSNRsqueezing}, we display the numerical results for the gain-to-instability ratio (GIR)  {to study the realistic impact of varying squeezing on stability of the gain in SNR. 
We keep the transmittance optimized to reach maximal mean displacement at the output as for the Fig.\ref{fig:SNRsqueezing}. A substantial difference can now be seen between the pyramidal, sequential and dynamical loop architectures in a contrast to fixed loop architectures. Indeed, while the pyramidal, sequential and dynamical loop architectures architecture has the GIR (blue points) that approximately scales as $~N^{1/2}$, the loop architecture with fixed optimized $t=t^*$ scales very differently as $~N^{1/4}$ (orange points) and, therefore, saturates much faster.
The loop architectures with a fixed transmittance $t=t_1=1-1/N$ (green points) have essentially saturating behaviour similar to strategy with fixed $t=t_2=1/2$ that was already inefficient in Fig.\ref{fig:SNRsqueezing}. The transmission optimization of the loop architecture with a single transmittance $t$ can moderately compromised beyond the mean displacement discussed here, between the mean SNR and the GIR. For an example, in Fig.\ref{fig:SNRSNRsqueezing} the loop strategy can reach $\mbox{GIR}^{(1000)}_{\mbox{SNR}}\approx 30$ at the cost of slight decrease of the mean SNR to $\mean{\mbox{SNR}_{out,sq}^{(1000)}}\approx 75$ and mean displacement to $13.4x_0$ for $N=1000$.}  

 {It is an experimentally testable example showing that such saturation of the GIR qualitatively limits the scaling-up reliability of the architecture under resource instability despite the SNR can behave rather similarly with the same power in the scaling. Hence, the difference between the different architectures must be accessed in terms of the new gain-to-instability ratio. The mean gain in the figure-of-merit is not sufficient to describe influence of resources {that are not stable} and can overestimate fixed loop architectures.}

 {This analysis of the impact of varying single-mode squeezing can be compared with the measurement-induced alternative that employs a combination of {two varying ortogonal squeezed states}. Each two-mode squeezing copy $i$ can be generated by mixing two orthogonally squeezed vacuum states with minimal variances $\sigma_{A,i}^2=\exp(-2r_0)\exp(-2\delta_{A,i})$ and $\sigma_{B,i}^2=\exp(-2r_0)\exp(-2\delta_{B,i})$ in pairs modes $A,i$ and $B,i$ at the balanced beamsplitter coupling. Measuring a value $x_m$ of one of the squeezed variables in the output mode of each copy, the conditional state has a displacement $x_{0,i}=x_m(\sigma_{A,i}^2-\sigma_{B,i}^{-2})/(\sigma_{A,i}^2+\sigma_{B,i}^{-2})$ and variance $\sigma_i^2=2\sigma_{A,i}^2\sigma_{B,i}^{-2}/(\sigma_{A,i}^2+\sigma_{B,i}^{-2})$ that appear in the other output mode of each copy. In the ideal limit of stable ($\delta_{A,i}=\delta_{B,i}=0$) and large ($r_0\gg 1$) squeezing $x_{0,i}\approx x_m$ and $\sigma_i^2\approx 2\exp(-2r_0)$. As the variance $\sigma_i^2$ of each copy is twice as large as in the previous single-mode cases, the measurement-induced preparation is less efficient. It already reduces ideally stable single-copy $\mbox{SNR}_{in}^{(1)}$ by factor $1/\sqrt{2}\approx 0.7$. Following the pyramidal, sequential, and dynamic loop architecture for {varying} squeezing, the reduction propagates and $\mbox{SNR}_{out}^{(N)}$ decreases by the same factor in this ideal limit. For stable squeezing parameter $r_0$, we reach ideally $\mbox{SNR}_{out}^{(N)}\approx\sqrt{N}x_0\sinh(2r_0)/\sqrt{\cosh(2r_0)}$. }  

Outside this idealistic limit, the varying squeezing parameters $\delta_{A,i}$ and $\delta_{B,i}$ affect also the mean displacements, however, $\mean{\mbox{SNR}_{out,sq}^{(N)}}$ does not significantly change for moderate squeezing instability $\sigma_r/r_0<1$. Therefore, $\mean{\mbox{SNR}_{out,sq}^{(N)}}$ reduces further as visibile from numerical simulation in Fig.\ref{fig:SNRsqueezing} (yellow points) by the rise in the standard deviation of the SNR's denominator. If the displacement is used for feedforward control of the output after the measurement of $x_m$ with a feedforward gain adjusted to minimise the variance of the output for the stable and necessarily known $r_0$, the nearly same $\mean{\mbox{SNR}_{out,sq}^{(N)}}$ is deterministically obtained for the pyramidal, sequential and dynamical loop architectures in the limit of moderate instability $\sigma_r/r_0<1$. Note that beyond this limit, known $r_0$ helps to deterministic feedforward strategy to achieve higher $\mean{\mbox{SNR}_{out,sq}^{(N)}}$. Despite the reduction of $\mean{\mbox{SNR}_{out,sq}^{(N)}}$ for both the measurement-induced strategies involving two-mode states, the GIR remains comparable for the pyramidal, sequential, and dynamical loop architectures to the single-mode strategies described by Eq.(\ref{eq:SNRSNRsqueezingPyr}) for a moderate instability $\sigma_r/r_0<1$, as shown in Fig.\ref{fig:SNRSNRsqueezing} (yellow points). As measurement-induced squeezing is extensively utilised as a resource for continuous-variable protocols, it demonstrates that moderate squeezing instability has a comparable impact to the direct squeezing discussed above.

Based on Eqs.(\ref{eq:SNRsqueezedstable},\ref{eq:SNRsqueezingapprox},\ref{eq:SNRsqueezingAsymptotic}), we can discuss in detail the effect of $\sigma_r$ on the mean SNR in Fig. \ref{fig:SNRvssigma}. The first observation is that the approximated expression for the mean SNR given by Eq.(\ref{eq:SNRsqueezingapprox}), depicted as a dotted blue line, tends to the asymptotic expression of the SNR in Eq.(\ref{eq:SNRsqueezingAsymptotic}) already for a number of inputs of order $N \approx 10$. A second observation is that as $N$ increases, the approximation of the SNR deviates from the ideal mean SNR more rapidly as $\sigma_r$ increases. A third point is that when considering the approximated mean SNR in Fig. \ref{fig:SNRvssigma}, one can observe that the SNR drops quickly when $\sigma_r$ increases. On the other hand, the numerically simulated mean SNR by averaging the formula (\ref{eq:SNRsqueezingfluctuation}) represented by the dots is higher and decays much slowly for larger $\sigma_r$ for all architectures, especially for larger $N$. A minor observation in Fig. \ref{fig:SNRvssigma} is that the fixed loop architecture provides a higher mean SNR for small $N$ in the ideal case than the pyramidal architecture. However, pyramidal architecture offers a higher SNR for larger $N$.

By taking the mean squeezing $r_0$ as a variable for the SNR in the different architecture, we see in  {Fig.\ref{fig:SNRvssqueezing}}, that they all scale exponentially up with $r_0$ for all $N$.  Similarly as in  {Fig. \ref{fig:SNRvssigma}} the fixed loop architecture has a higher SNR for small number of modes $N$ while the pyramidal, sequential and dynamical loop architectures provide a higher SNR for larger number of input modes $N$. Moreover, the larger the number of input states in  {Fig. \ref{fig:SNRvssigma}}, the larger becomes the difference in SNR between the different architectures. This shows that a larger number of {varying squeezings} in input states makes the discrimination between different architectures more visible. However, the relative difference between the architecture seems to be constant. Hence, the different scaling for each architecture is impacted by the fluctuation $\sigma_r$ rather than the mean squeezing $r_0$.

 {The Fig.\ref{fig:SNRSNRvssigma} shows that the GIR decreases with the fluctuations in squeezing $\sigma_r$ according to its approximations in Eqs.(\ref{eq:SNRSNRsqueezingPyr},\ref{eq:SNRSNRsqueezingPyrAsympt}). On one hand, in the limit of small flucuations $\sigma_r \ll 1$, the  GIR is diverging to infinity. On the other hand, in the limit of largely varying squeezings, the GIR of the different architectures and different number of input modes converges to values smaller than $10$. In this regime of large fluctuations, the number of modes do not play as significant role in discriminating the performances of the different protocols. Hence, one can conclude that the interest of using GIR resides in the regime of moderate to low variations of the squeezings in input states where its discriminating power between between the different architecture at different scales is the largest. As a final remark for Fig.\ref{fig:SNRSNRvssigma} , the GIR typically increase with the number of modes (see for example the lines in orange and blue). However, in the case of the architecture using a beam-spitter with a fixed transmittance of $(N-1)/N$ (green lines), the GIR do not vary much with the number of modes N. This is the signature of a saturation in the GIR as it can be observed in Fig.\ref{fig:SNRvssqueezing}.}

 {Finally, Fig.\ref{fig:SNRSNRvssqueezing}  allows us to conclude that the GIR is independent of the mean squeezing of the input states $r_0$. Hence, the GIR is only determined by the number of input states and the varying squeezing of these states as expected. Therefore, the GIR can be used independently to the squeezing regime.}

\section{Conclusions}

We provide a methodology to analyse instability of resources used in multicopy bosonic protocols and applications by introducing gain-to-instability ratio. 

Already for Gaussian states we demonstrated an essential difference in the scaling properties of the architectures.  The average gains represented by a signal-to-noise ratio for the displacement might be very similar, but for the pyramidal, sequential and loop architecture architecture, the  gain-to-instability ratio grows as $~N^{1/2}$ whereas the optimized loop architecture grows as $~N^{1/4}$ , although the gains themself scale analogically as $~N^{1/2}$. The other, more simpler and fixed, architectures are quickly saturating over $N$, or even worse, the gain-to-instability ratio decreases and remains below one. The effect of losses reduces the SNR, that stops to increase for more than $50 \%$ of loss, but at the same time, it increase the stability of the output noise leading to a larger GIR (see Appendix \ref{app:losses}). This demonstrates this new quantifier's relevance in predicting truly scalable architectures  for even slightly different resources. For a comparison, using linear measurements, one could enhance the performances of the architecture by feed-forward strategies \cite{glockl_squeezed-state_2006} or by modifying the type of average at the constructive output port \cite{lassen_experimental_2010}  {(see Appendix \ref{sec:harmonicmean}).} However, this potential might be at the cost of  even more sophisticated optimization respective to prior knowledge about the input states or a low success rate.
\newline
The next step is verifying such scalings experimentally and then extending the theory analysis with experimental tests much more broadly. The current time-multiplexed interference optical experiments  \cite{yonezu_time-domain_2023,enomoto_programmable_2021,takeda_-demand_2019} with squeezed light are ideal platforms for pioneering proof-of-principle experimental tests with more than a hundred copies needed to predict new scaling of the gain-to-instability ratio.
\newline
Subsequently, it is relevant to consider the scalability of Gaussian entangled states and the resources that are needed for large structures of entanglement like cluster states  \cite{van_loock_detecting_2003,gu_quantum_2009,gonzalez-arciniegas_cluster_2021}. 
Finally, the scalability of quantum non-Gaussian resources  \cite{zapletal_experimental_2021} can be predicted theoretically and verified after they become available with sufficient experimental rates for interference experiments. Such diverse investigations can bring the necessary knowledge to operationally navigate experiments towards more scalable and reliable architectures in even slightly unstable conditions.  {Once these scalable interferences are experimentally tested, the analysis of the figures of merit in applications of squeezed states will ensue, enabling valuable comparisons with the scalability of SNR.} 

\section{Acknolwedgment}

M.A. acknowledges the framework of the Contrat de Plan
Etat-Region (CPER) WaveTech@HdF supported by Ministry of Higher
Education and Research, Région Hauts-de-France (HdF), Lille Metropole
(MEL) and the Institute of Physics (INP) of National Centre for
Scientific Research (CNRS) and European Union’s HORIZON Research and Innovation Actions under Grant Agreement no. 101080173 (CLUSTEC). R.F. acknowledges the Quantera project CLUSSTAR (8C24003) of the Czech Ministry of Education, Youth and Sport and EU under Grant Agreements No. 101017733 and 731473 {and also the project No. LUC25006 of MEYS Czech Republic.}   

\section{SUPPLEMENTARY MATERIAL}
\label{sec:supplementary}

\begin{figure*}[!ht]
	\centering
	
\includegraphics[width=0.49\textwidth]{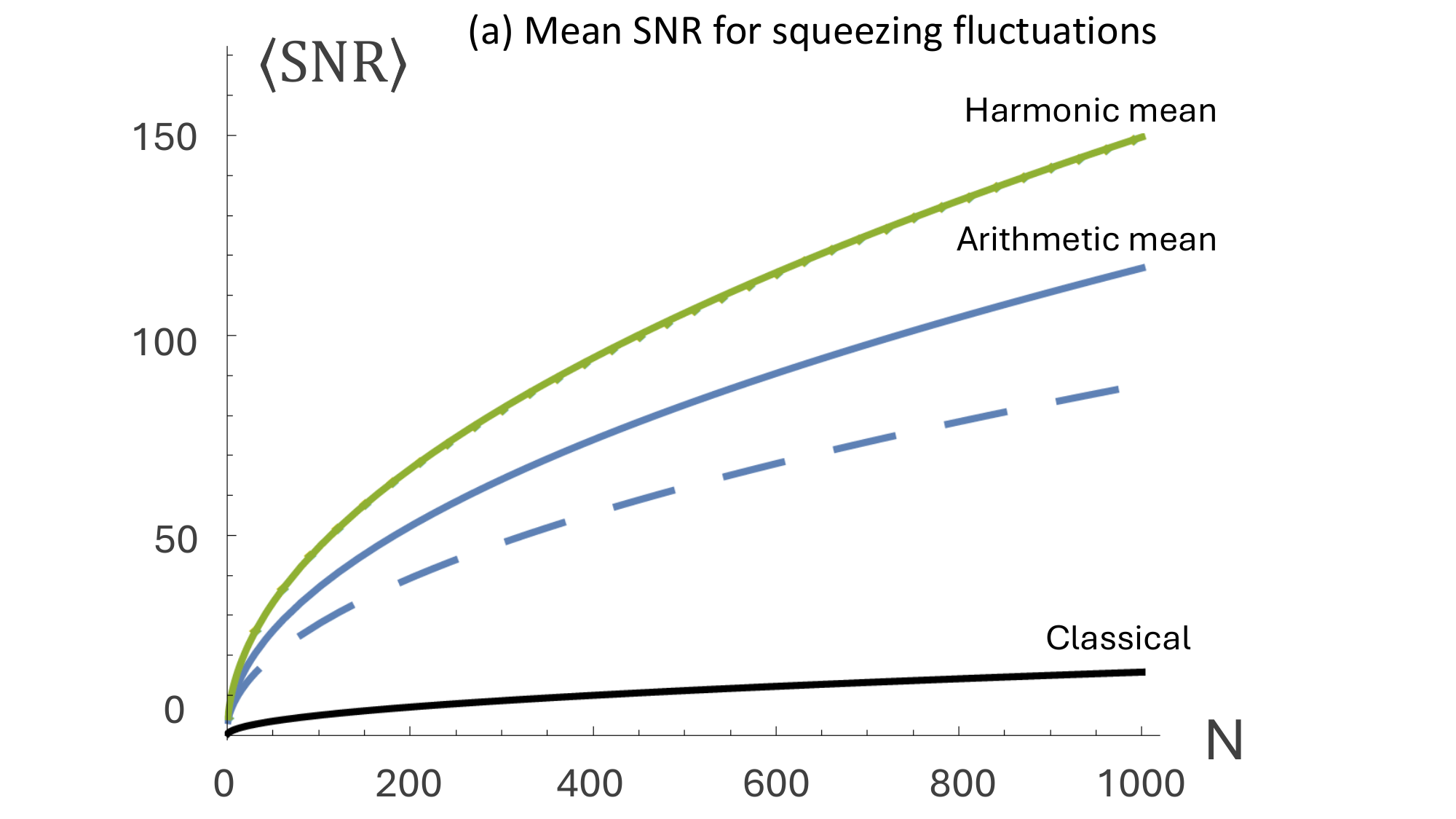}\label{fig:SNRsqueezingHarm}
\includegraphics[width=0.49\textwidth]{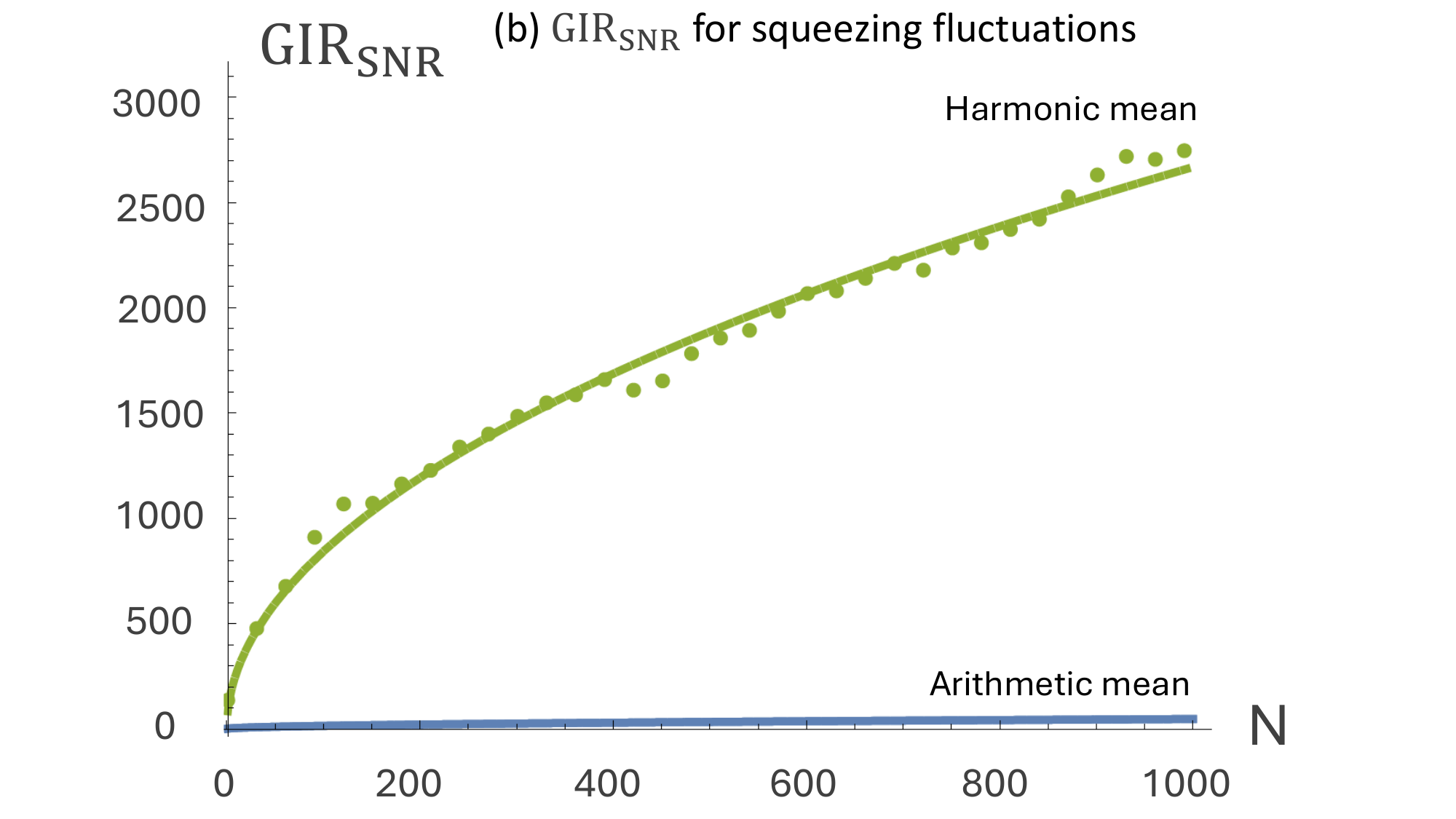} \label{fig:SNRSNRsqueezingHarm}
	\caption{By changing the protocol allowing for conditional measurement on all output modes but the constructive modes, on can change further enhance the scaling capability. Indeed, not only does the SNR slightly improves but, more importantly, the GIR improve by order of magnitudes. This confirms the predictions that this harmonic averaging of the second order moment enhance the stability of the constructive output squeezed state. In green is the harmonic mean and in blue the deterministic protocol.}
\end{figure*}

\begin{figure*}[!ht]
	\centering
	\begin{subfigure}{
\includegraphics[height=0.13\textheight,keepaspectratio]{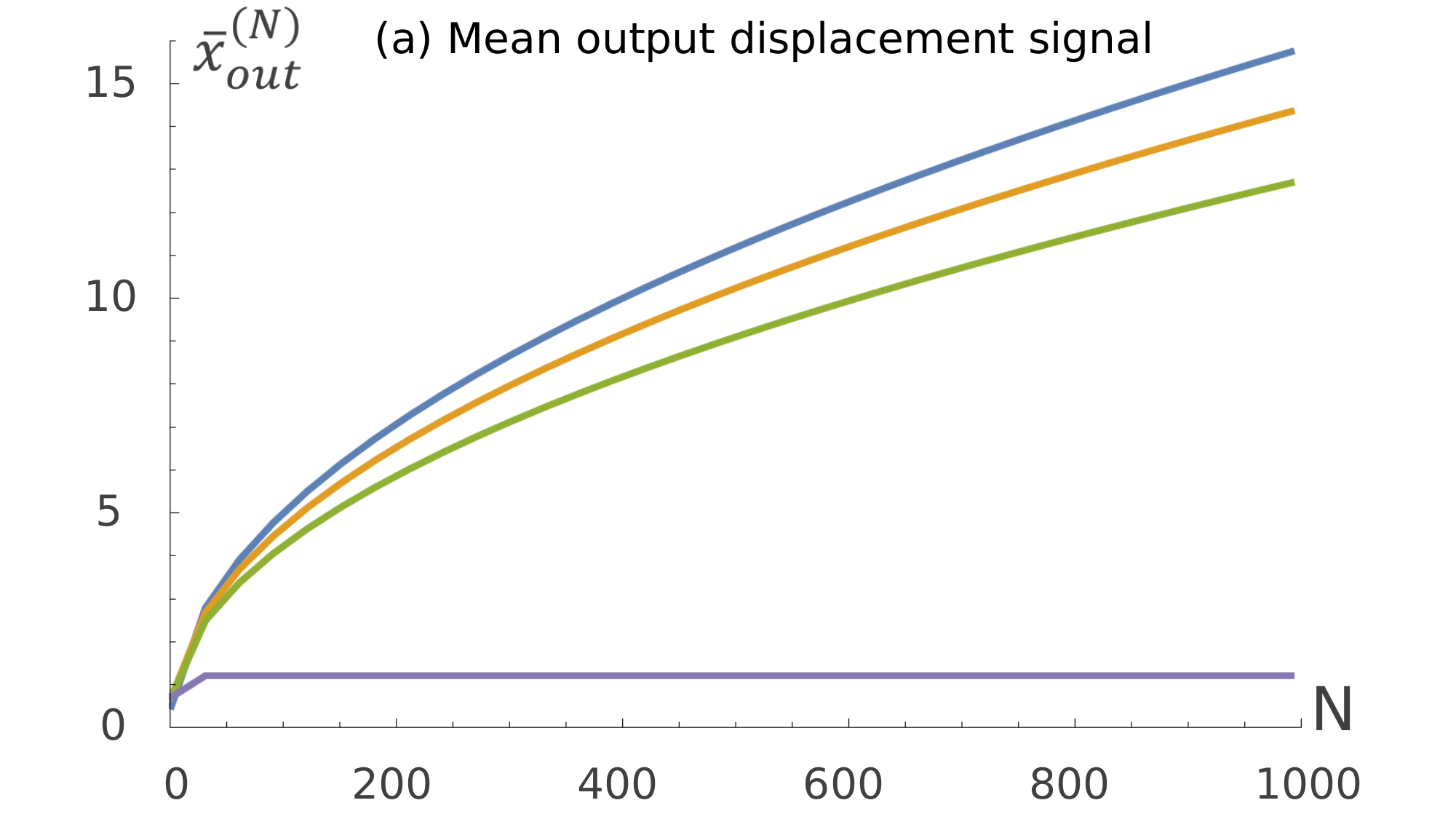}\label{fig:signalplot}}
\end{subfigure}
\begin{subfigure}{
\includegraphics[height=0.13\textheight,keepaspectratio]{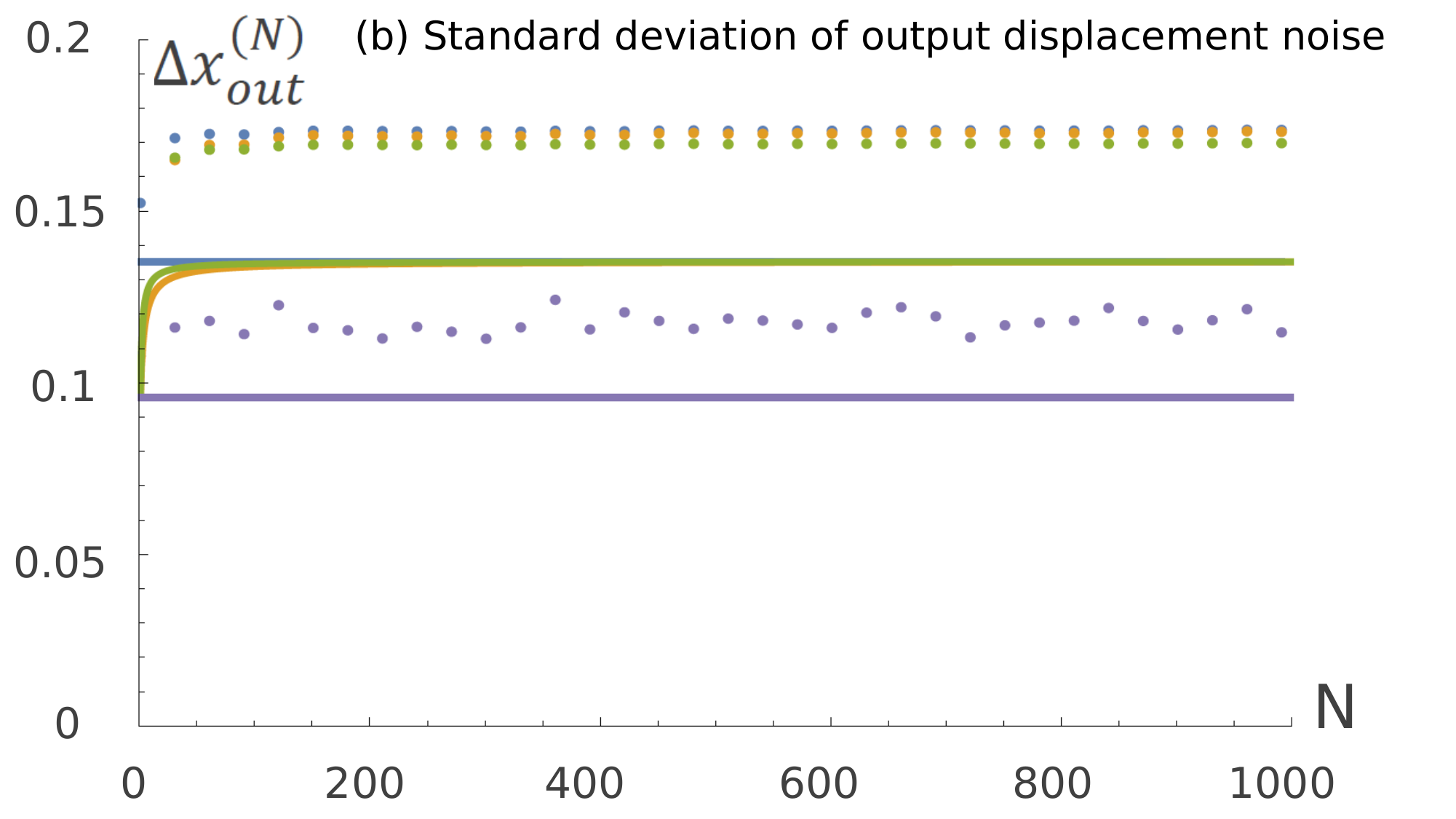}\label{fig:noiseplot}}
\end{subfigure}
\begin{subfigure}{
	\includegraphics[height=0.13\textheight,keepaspectratio]{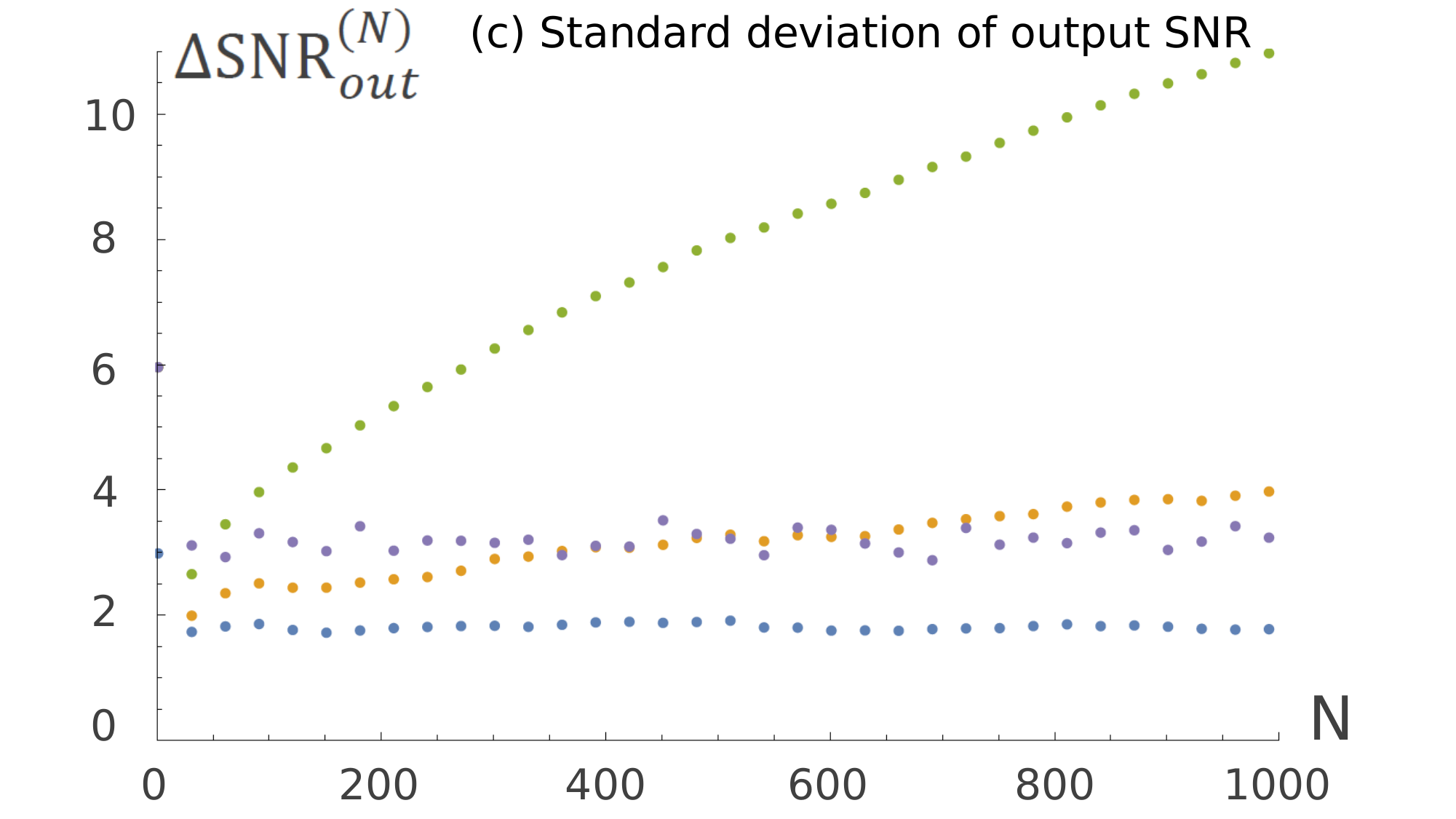}\label{fig:SNRSTDplot}}
	  \end{subfigure}
	\caption{The different quantities involved in the numerical evaluation of $\mbox{SNR}$ and $\mbox{GIR}$ are displayed : (a) is the plot of the mean signal, (b) is the plot of the noise, (c) is the plot of the standard deviation of the $\mbox{SNR}$. Figures show that the scaling of the $\mbox{SNR}$ comes from the signal and that the accumulation of the noise in the interferometer is not sufficient to compromise the scaling. However, the standard deviation of the $\mbox{SNR}$ is critical in the $\mbox{GIR}$ scaling. Remarkably, only a slightly increase of the standard deviation of the $\mbox{SNR}$in the pyramidal, sequential and dynamical loop architecture (orange dots) is responsible the $~N^{1/4}$ scaling for the $\mbox{GIR}$.}
\end{figure*}

\subsection{Harmonic mean strategy}\label{sec:harmonicmean}

A possible solution to enhance the interference effects is to allow for a protocol that is implementing an harmonic mean of the quadrature variance instead of the arithmetic mean of the pyramidal, sequential and dynamical loop protocols. Such protocol has been proposed and experimentally tested in quantum optics \cite{lassen_experimental_2010}. In this harmonic mean protocol, the $N$ input states interfere on an beam-splitter architecture similar to the pyramidal architecture of Fig.2a, the amplitude quadratures of the $N-1$ outputs are measured and conditioned to be arbitrarily close to zero. The constructive output state has a displacement equal to $\sqrt{N} x_0$ of the combined initial displacement $x_0$ of the input states and a variance that is, by construction, equal to $(\Delta x_{out}^2)^{-1} = 1/N \sum_i (\Delta x_{i}^2)^{-1}$. Hence, for amplitude quadrature squeezed input states with individual squeezing $r_i = r_0 + \delta r_i$, the output SNR writes
\begin{equation}
	\mbox{SNR}_{out,harm}^{(N)} = x_0 e^r_0 \sqrt{\sum_i^N e^{+2 \delta r_i}}.
\end{equation}
By numerical evaluation, this protocol provides,  a multiplicative enhancement over the averaging SNR for the passive pyramidal protocol and maintain a $\sqrt{N}$ scaling (see Fig.5a). However, it performs far better in the GIR as it maintain the $\sqrt{N}$ scaling but with a multiplicative factor 2 orders of magnitude larger compared to the passive pyramidal  architecture presented in the main text (see Fig.5b). Hence, as the costs of this conditional harmonic protocol does not show significantly in the gain, its advantage over the deterministic pyramical, sequential and dynamical loop protocols are clear in the more advance statistical evaluation provided by the GIR. 

\subsection{Fixed loop strategy analysis}

{In the fixed loop architecture, the scaling law is hidden in the fraction appearing on the right hand side of Eq.(\ref{eq:SNRSqueezingLoop}). As this term strongly depends on the value of the transmittance $t$ of the beam splitter, one should optimize the numerator of Eq.(\ref{eq:SNRSqueezingLoop}) in order to maximize the value of the mean displacement. The transmitance $t=t^*$ is a function of the number of input modes $N$ and can be found as the roots of a polynomial function in of order $N$. The numerical simulations for displaying the $\mean{\mbox{SNR}}$ and the GIR$=\mean{\mbox{SNR}}/\Delta \mbox{SNR}$ are the orange curves in Fig.3. The behaviour of the mean SNR follows a similar scaling law proportional to $N^{1/2}$ as for the pyramidal, sequential and dynamical loop architectures above. However, the GIR is closer to a scaling law of order $N^{1/4}$. As visible on Fig.\ref{fig:noiseplot}, it is the standard deviation $\Delta \mbox{SNR}$ that is critical in the limitation of the scaling for the loop architecture. Indeed, even a slight increase in the $\Delta \mbox{SNR}$ can seriously limit the scaling laws in the $\mbox{GIR}$, even if $\Delta \mbox{SNR}$ (in orange) is only twice larger than for the pyramidal, sequential and dynamical loop architectures above (in blue).  This supports the relevance of the $\mbox{GIR}$ to assess critical scaling behaviours.}

\subsection{Effect of losses on the scalability}\label{app:losses}

As our main paper aim to emphasize the role played by the noise instability in the scaling laws, we considered the case of a large number of input pure states interfere on interferometers. However, losses are a critical and unavoidable feature in all optics experiments and are expected to limit, possibly even more than the noise instability, the performance of the protocols. Hence, we consider here the effect of such losses by introducing a pure loss channel, where the input state is interacting with the vacuum through a beam splitter of transmittance $\eta$, at the input of  \textit{each} beam splitter in the pyramidal architecture. We evaluate the output signal, noise, SNR and GIR of our protocols for different values of  $\eta$. As expected, the losses severely reduce the scalability in the SNR (see Fig.7). However, it is important for an experimental verification that for  $\eta>0.5$, the scaling law is conserved and that only the prefactor is reduced. This behaviour is expected as the loss channel reduces both the signal and the noise. Simultaneously, as the rise of the SNR is worse, the GIR is smaller for larger values of $\eta$ (see Fig.7). Indeed, the losses are stabilizing the protocol closer to classical limit and as mentioned in the paper, the GIR becomes very large for classical stable states.
\begin{figure*}[!ht]
	\centering
\begin{subfigure}{	
\includegraphics[width=0.45\textwidth]{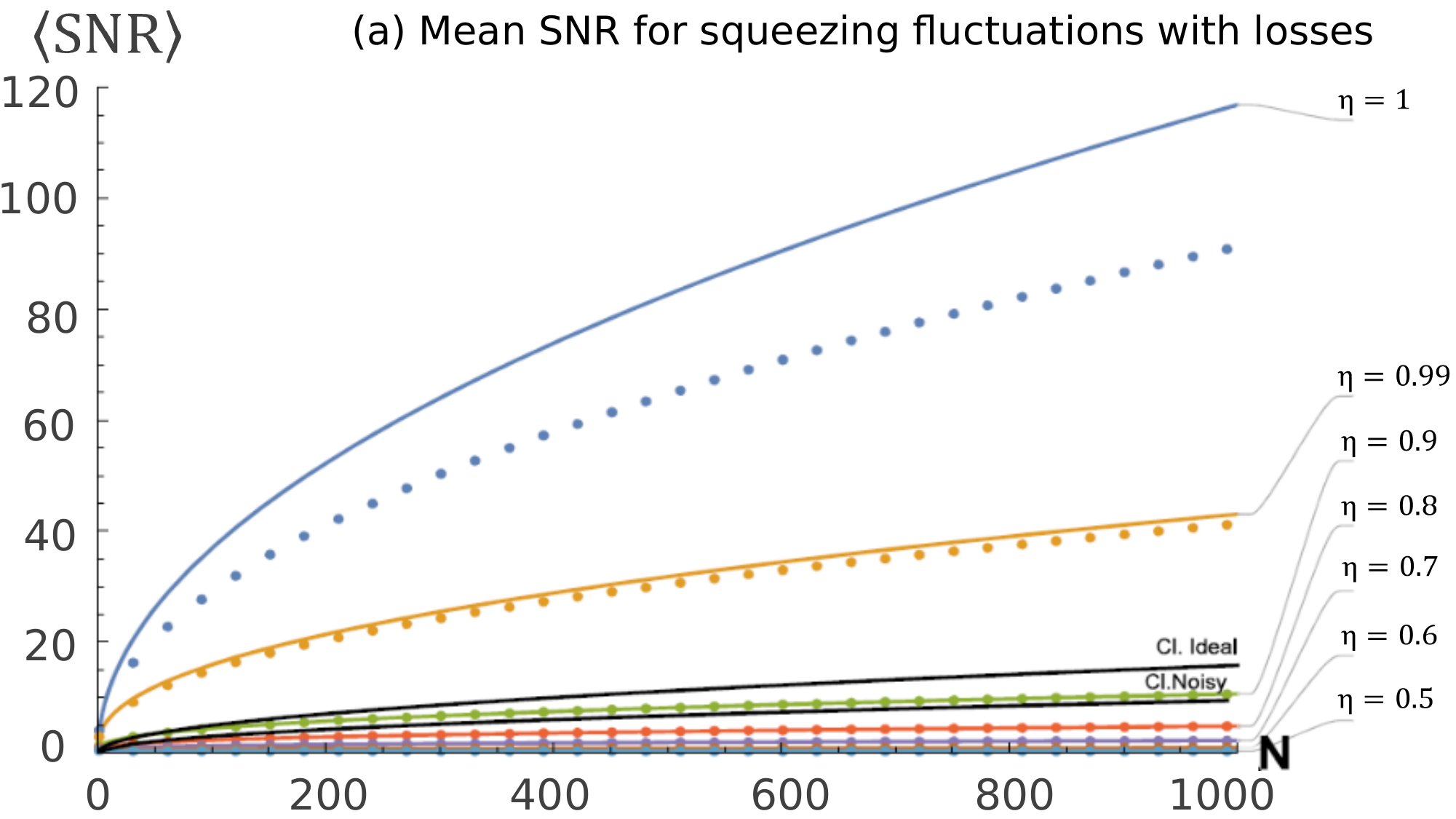}\label{fig:SNRsqueezingLosses}}
\end{subfigure}
\begin{subfigure}{
\includegraphics[width=0.45\textwidth]{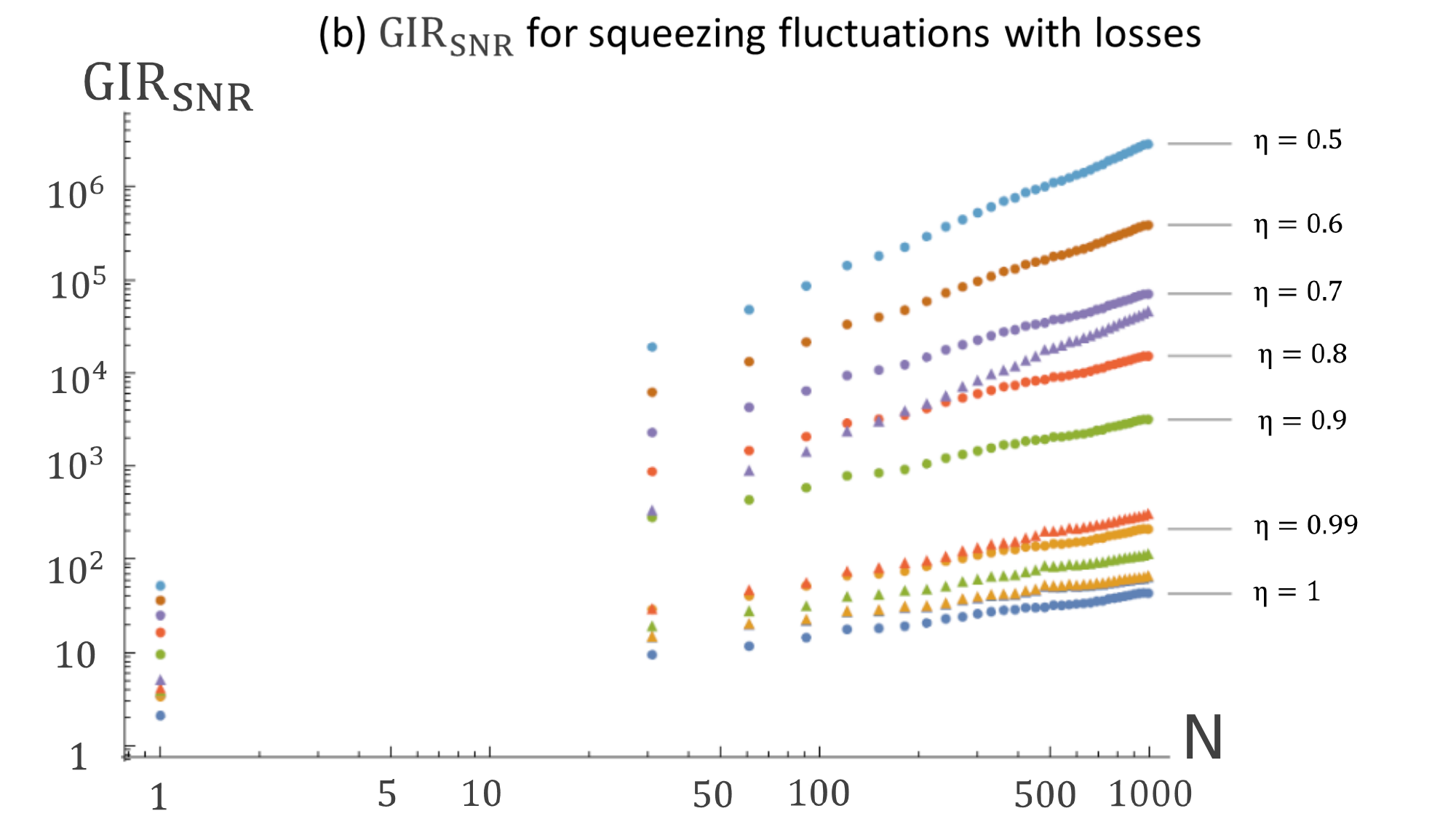} \label{fig:GIRsqueezingLosses}}
\end{subfigure}
	\caption{SNR and GIR for the pyramidal architecture with pure loss channel added in the input of each beam splitter. The round dots are numerical simulations for the squeezed states and the triangles are the numerical simulation of the classical states. The addition of pure loss on the input of each beam-splitter in the dynamical, sequential and dynamical loop architecture, significantly reduces the SNR. For a transmittivity of about $\eta=0.8$, the SNR is reduced to the classically attainable SNR. Hence, losses makes the SNR to loose their squeezing advantage even for relatively small losses. On the other hand, it stabilizes the noise and hence makes the GIR to be significantly higher than the loss-less case.}
\end{figure*}

\subsection{Fit parameter values}\label{appendixD:fitparametervalues}
\begin{table}[h!]
	\begin{center}
		\begin{tabular}{| c | c | c | c || c | c |}
			\hline
			Architecture & $x_0$ & $r_0$ & $\sigma_r$ & $a$ & $b$ \\
			\hline
			Pyramidal & $2$ & $2$ & $0.125$ & $15$ & $0.50$ \\ 
			\hline
			Loop & $2$ & $2$ & $0.125$ & $15$ & $0.49$  \\
			\hline
			Loop (N-1)/N & $2$ & $2$ & $0.125$ & $14$ & $0.48$ \\  
			\hline
			\hline
			Pyramidal & $0.5$ & $2$ & $0.5$ & $2.9$ & $0.50$ \\ 
			\hline
			Loop & $0.5$ & $2$ & $0.5$ & $3.0$ & $0.48$  \\
			\hline
			Loop (N-1)/N & $0.5$ & $2$ & $0.5$ & $2.9$ & $0.47$ \\  
			\hline
			\hline
			Pyramidal & $0.5$ & $2$ & $0.25$ & $3.5$ & $0.50$ \\ 
			\hline
			Loop & $0.5$ & $2$ & $0.25$ & $3.5$ & $0.49$  \\
			\hline
			Loop (N-1)/N & $0.5$ & $2$ & $0.25$ & $3.3$ & $0.48$ \\  
			\hline
			\hline
			Pyramidal & $0.5$ & $1$ & $0.5$ & $1.1$ & $0.50$ \\ 
			\hline
			Loop & $0.5$ & $1$ & $0.5$ & $1.1$ & $0.49$ \\
			\hline
			Loop (N-1)/N & $0.5$ & $1$ & $0.5$ & $1.0$ & $0.48$ \\  
			\hline
			\hline
			Pyramidal & $0.1$ & $1$ & $0.5$ & $0.217$ & $0.496$ \\ 
			\hline
			Loop & $0.1$ & $1$ & $0.5$ & $0.220$ & $0.481$ \\
			\hline
			Loop (N-1)/N & $0.1$ & $1$ & $0.5$ & $0.213$ & $0.474$ \\  
			\hline
		\end{tabular}
		\caption{Exponential Fit function $a N^b$ and fitting parameters for $\mean{\mbox{SNR}}$.}\label{tab:SNRfit}
	\end{center}
\end{table}

\begin{table}[h!]
	\begin{center}
		\begin{tabular}{| c | c | c | c || c | c |}
			\hline
			Architecture & $x_0$ & $r_0$ & $\sigma_r$ & $a$ & $b$ \\
			\hline
			\hline
			Pyramidal & $2$ & $2$ & $0.125$ & $9.8$ & $0.46$ \\ 
			\hline
			Loop & $2$ & $2$ & $0.125$ & $18$ & $0.25$ \\
			\hline
			Loop (N-1)/N & $2$ & $2$ & $0.125$ & $19$ & $0.038$ \\
			\hline
			\hline
			Pyramidal & $0.5$ & $2$ & $0.5$ & $1.5$ & $0.51$ \\ 
			\hline
			Loop & $0.5$ & $2$ & $0.5$ & $3.5$ & $0.27$ \\
			\hline
			Loop (N-1)/N & $0.5$ & $2$ & $0.5$ & $3.7$ & $0.10$ \\  
			\hline
			\hline
			Pyramidal & $0.5$ & $2$ & $0.25$ & $2.7$ & $0.57$ \\ 
			\hline
			Loop & $0.5$ & $2$ & $0.25$ & $8.8$ & $0.23$ \\
			\hline
			Loop (N-1)/N & $0.5$ & $2$ & $0.25$ & $8.4$ & $0.043$ \\  
			\hline
			\hline
			Pyramidal & $0.5$ & $1$ & $0.5$ & $1.3$ & $0.53$ \\ 
			\hline
			Loop & $0.5$ & $1$ & $0.5$ & $4.0$ & $0.21$ \\
			\hline
			Loop (N-1)/N & $0.5$ & $1$ & $0.5$ & $4.3$ & $0.050$ \\  
			\hline
			\hline
			Pyramidal & $0.1$ & $1$ & $0.5$ & $1.9$ & $0.47$\\ 
			\hline
			Loop & $0.1$ & $1$ & $0.5$ & $3.5$ & $0.27$ \\
			\hline
			Loop (N-1)/N & $0.1$ & $1$ & $0.5$ & $4.9$ & $0.054$ \\  
			\hline
		\end{tabular}
		\caption{Exponential Fit function $a N^b$ and fitting parameters for $\mbox{GIR}=\langle \mbox{SNR} \rangle/\Delta \mbox{SNR}$.}\label{tab:GIRfit}
	\end{center}
\end{table}


\begin{thebibliography}{99}

\bibitem{cerf_quantum_2007} Nicolas J. Cerf, Gerd Leuchs, and Eugeen S. Polzik. Quantum Information with Continuous Variables of Atoms
and Light. Imperial College Press, 2007.

\bibitem{weedbrook_gaussian_2012-1}
Christian Weedbrook, Stefano Pirandola, Raul Garc\'{ı}a-Patr\' on, Nicolas J. Cerf, Timothy C. Ralph, Jeff H. Shapiro, and Seth Lloyd. Gaussian
quantum information. Reviews of Modern Physics,
84(2):621–669, May 2012. 

\bibitem{walschaers_non-gaussian_2021-1}
Mattia Walschaers. Non-Gaussian Quantum States and
Where to Find Them. PRX Quantum, 2(3):030204,
September 2021.

\bibitem{rakhubovsky_quantum_2024}
Andrey A. Rakhubovsky, Darren W. Moore, and Radim
Filip. Quantum non-Gaussian optomechanics and
electromechanics. Progress in Quantum Electronics,
93:100495, January 2024.

\bibitem{lachman_quantum_2022} 
Luk\' a\v s Lachman and Radim Filip. Quantum non-
Gaussianity of light and atoms. Progress in Quantum
Electronics, 83:100395, May 2022.

\bibitem{huang_optical_2015}
K. Huang, H. Le Jeannic, J. Ruaudel, V.B. Verma, M.D. Shaw, F. Marsili, S.W. Nam, E. Wu, H. Zeng, Y.-C. Jeong, R. Filip, O. Morin and J. Laurat, Optical Synthesis of Large-Amplitude Squeezed Coherent-State Superpositions with Minimal Resources, PRL 115, 023602 (2015)

\bibitem{hacker_deterministic_2019}
Hacker et al.
Deterministic creation of entangled atom–light Schrödinger-cat states. Nature Photon
13, 110–115 (2019)

\bibitem{simon_experimental_2024}
Simon et al.
Experimental Demonstration of a Versatile and Scalable Scheme for Iterative Generation of Non-Gaussian States of Light
Phys. Rev. Lett. 133, 173603 (2024)

\bibitem{gottesman_encoding_2001}
Daniel Gottesman, Alexei Kitaev, and John Preskill. Encoding
a qubit in an oscillator. Physical Review A,
64(1):012310, June 2001.

\bibitem{marek_general_2018}
Petr Marek, Radim Filip, Hisashi Ogawa, Atsushi Sakaguchi,
Shuntaro Takeda, Jun-ichi Yoshikawa, and Akira
Furusawa. General implementation of arbitrary nonlinear
quadrature phase gates. Physical Review A,
97(2):022329, February 2018.

\bibitem{bravyi_universal_2005}
Sergey Bravyi and Alexei Kitaev. Universal quantum
computation with ideal Clifford gates and noisy ancillas.
Physical Review A, 71(2):022316, February 2005.

\bibitem{walschaers_tailoring_2018}
Mattia Walschaers, Supratik Sarkar, Valentina Parigi,
and Nicolas Treps. Tailoring Non-Gaussian Continuous-
Variable Graph States. Physical Review Letters, 121(22):220501, November 2018.

\bibitem{walschaers_emergent_2023}
Mattia Walschaers, Bhuvanesh Sundar, Nicolas Treps,
Lincoln D Carr, and Valentina Parigi. Emergent complex
quantum networks in continuous-variables non-Gaussian
states. Quantum Science and Technology, 8(3):035009,
July 2023.


\bibitem{lewis-swan_robust_2018}
Robert J. Lewis-Swan, Matthew A. Norcia, Julia R. K.
Cline, James K. Thompson, and Ana Maria Rey. Robust
Spin Squeezing via Photon-Mediated Interactions
on an Optical Clock Transition. Physical Review Letters,
121(7):070403, August 2018.

\bibitem{schulte_prospects_2020} 
Marius Schulte, Christian Lisdat, Piet O. Schmidt, Uwe
Sterr, and Klemens Hammerer. Prospects and challenges
for squeezing-enhanced optical atomic clocks. Nature
Communications, 11(1):5955, November 2020.

\bibitem{pedrozo-penafiel_entanglement_2020}
Edwin Pedrozo-Pe\~ nafiel, Simone Colombo, Chi Shu, Albert
F. Adiyatullin, Zeyang Li, Enrique Mendez, Boris
Braverman, Akio Kawasaki, Daisuke Akamatsu, Yanhong
Xiao, and Vladan Vuleti\' c. Entanglement on an
optical atomic-clock transition. Nature, 588(7838):414–
418, December 2020.


\bibitem{robinson_direct_2024}
John M. Robinson, Maya Miklos, Yee Ming Tso, Colin J.
Kennedy, Tobias Bothwell, Dhruv Kedar, James K.
Thompson, and Jun Ye. Direct comparison of two
spin-squeezed optical clock ensembles at the 10-17 level.
Nature Physics, January 2024.

\bibitem{degen_quantum_2017}
C. L. Degen, F. Reinhard, and P. Cappellaro. Quantum
sensing. Rev. Mod. Phys., 89(3):035002, July 2017.


\bibitem{stokowski_integrated_2023}
Hubert S. Stokowski, Timothy P. McKenna, Taewon
Park, Alexander Y. Hwang, Devin J. Dean, Oguz Tolga
Celik, Vahid Ansari, Martin M. Fejer, and Amir H.
Safavi-Naeini. Integrated quantum optical phase sensor
in thin film lithium niobate. Nature Communications,
14(1):3355, June 2023.


\bibitem{malia_distributed_2022}
Benjamin K. Malia, YunfanWu, Juli\'{a}n Mart\'{ı}nez-Rinc\'{o}n,
and Mark A. Kasevich. Distributed quantum sensing
with mode-entangled spin-squeezed atomic states.
Nature, 612(7941):661–665, December 2022.


\bibitem{xia_entanglement-enhanced_2023}
Yi Xia, Aman R. Agrawal, Christian M. Pluchar, Anthony
J. Brady, Zhen Liu, Quntao Zhuang, Dalziel J.
Wilson, and Zheshen Zhang. Entanglement-enhanced optomechanical
sensing. Nature Photonics, 17(6):470–477,
June 2023.


\bibitem{franke_quantum-enhanced_2023}
Johannes Franke, Sean R. Muleady, Raphael Kaubruegger,
Florian Kranzl, Rainer Blatt, Ana Maria Rey,
Manoj K. Joshi, and Christian F. Roos. Quantumenhanced
sensing on optical transitions through finiterange
interactions. Nature, 621(7980):740–745, September
2023.


\bibitem{nielsen_deterministic_2023}
Jens A. H. Nielsen, Jonas S. Neergaard-Nielsen, Tobias
Gehring, and Ulrik L. Andersen. Deterministic Quantum
Phase Estimation beyond N00N States. Physical Review
Letters, 130(12):123603, March 2023.


\bibitem{madsen_continuous_2012}
Lars S. Madsen, Vladyslav C. Usenko, Mikael Lassen,
Radim Filip, and Ulrik L. Andersen. Continuous variable
quantum key distribution with modulated entangled
states. Nature Communications, 3(1):1083, September
2012.


\bibitem{gehring_implementation_2015}
Tobias Gehring, Vitus H\"{a}ndchen, J\"{o}rg Duhme, Fabian
Furrer, Torsten Franz, Christoph Pacher, Reinhard F.
Werner, and Roman Schnabel. Implementation of
continuous-variable quantum key distribution with
composable and one-sided-device-independent security
against coherent attacks. Nature Communications,
6(1):8795, October 2015.


\bibitem{jacobsen_complete_2018} 
Christian S. Jacobsen, Lars S. Madsen, Vladyslav C.
Usenko, Radim Filip, and Ulrik L. Andersen. Complete
elimination of information leakage in continuous-variable
quantum communication channels. npj Quantum
Information, 4(1):32, July 2018.


\bibitem{kovalenko_frequency-multiplexed_2021}
Olena Kovalenko, Young-Sik Ra, Yin Cai, Vladyslav C.
Usenko, Claude Fabre, Nicolas Treps, and Radim Filip.
Frequency-multiplexed entanglement for continuousvariable
quantum key distribution. Photonics Research,
9(12):2351, December 2021.


\bibitem{suleiman_40_2022}
I Suleiman, J A H Nielsen, X Guo, N Jain, J Neergaard-
Nielsen, T Gehring, and U L Andersen. 40 km fiber transmission
of squeezed light measured with a real local oscillator.
Quantum Science and Technology, 7(4):045003,
October 2022.

\bibitem{zhong_quantum_2020}
Han-Sen Zhong, Hui Wang, Yu-Hao Deng, Ming-Cheng
Chen, Li-Chao Peng, Yi-Han Luo, Jian Qin, Dian Wu,
Xing Ding, Yi Hu, Peng Hu, Xiao-Yan Yang, Wei-Jun
Zhang, Hao Li, Yuxuan Li, Xiao Jiang, Lin Gan, Guangwen
Yang, Lixing You, Zhen Wang, Li Li, Nai-Le Liu,
Chao-Yang Lu, and Jian-Wei Pan. Quantum computational
advantage using photons. Science, 370(6523):1460–
1463, December 2020.


\bibitem{wang_boson_nodate}
Hui Wang, Jian Qin, Xing Ding, Ming-Cheng Chen,
Si Chen, Xiang You, Yu-Ming He, Xiao Jiang, L You,
J J Renema, Sven H\"{o}fling, Chao-Yang Lu, and Jian-Wei
Pan. Boson sampling with 20 input photons in 60- mode
interferometers at 1014 state spaces. Physical Review Letters 123(25):250503, December, 2019


\bibitem{zhong_phase-programmable_2021}
Han-Sen Zhong, Yu-Hao Deng, Jian Qin, Hui Wang,
Ming-Cheng Chen, Li-Chao Peng, Yi-Han Luo, DianWu,
Si-Qiu Gong, Hao Su, Yi Hu, Peng Hu, Xiao-Yan Yang,
Wei-Jun Zhang, Hao Li, Yuxuan Li, Xiao Jiang, Lin
Gan, Guangwen Yang, Lixing You, Zhen Wang, Li Li,
Nai-Le Liu, Jelmer J. Renema, Chao-Yang Lu, and Jian-
Wei Pan. Phase-Programmable Gaussian Boson Sampling
Using Stimulated Squeezed Light. Physical Review
Letters, 127(18):180502, October 2021.


\bibitem{asavanant_generation_2019}
Warit Asavanant, Yu Shiozawa, Shota Yokoyama,
Baramee Charoensombutamon, Hiroki Emura, Rafael N.
Alexander, Shuntaro Takeda, Jun-ichi Yoshikawa, Nicolas
C. Menicucci, Hidehiro Yonezawa, and Akira Furusawa.
Generation of time-domain-multiplexed twodimensional
cluster state. Science, 366(6463):373–376,
October 2019.


\bibitem{larsen_deterministic_2019} 
Mikkel V. Larsen, Xueshi Guo, Casper R. Breum,
Jonas S. Neergaard-Nielsen, and Ulrik L. Andersen.
Deterministic generation of a two-dimensional cluster
state. Science, 366(6463):369–372, October 2019.


\bibitem{konno_logical_2024} 
Shunya Konno, Warit Asavanant, Fumiya Hanamura,
Hironari Nagayoshi, Kosuke Fukui, Atsushi Sakaguchi,
Ryuhoh Ide, Fumihiro China, Masahiro Yabuno, Shigehito Miki, Hirotaka Terai, Kan Takase, Mamoru Endo,
Petr Marek, Radim Filip, Peter Van Loock, and Akira
Furusawa. Logical states for fault-tolerant quantum computation
with propagating light. Science, 383(6680):289–
293, January 2024.

\bibitem{larsen_integrated_2025}
Larsen, M.V., Bourassa, J.E., Kocsis, S. et al. Integrated photonic source of Gottesman–Kitaev–Preskill qubits. Nature 642, 587–591 (2025).

\bibitem{miwa_exploring_2014}
Yoshichika Miwa, Jun-ichi Yoshikawa, Noriaki Iwata,
Mamoru Endo, Petr Marek, Radim Filip, Peter
Van Loock, and Akira Furusawa. Exploring a New
Regime for Processing Optical Qubits: Squeezing and
Unsqueezing Single Photons. Physical Review Letters,
113(1):013601, July 2014.

\bibitem{larsen_deterministic_2021}
Mikkel V. Larsen, Xueshi Guo, Casper R. Breum,
Jonas S. Neergaard-Nielsen, and Ulrik L. Andersen.
Deterministic multi-mode gates on a scalable photonic
quantum computing platform. Nature Physics,
17(9):1018–1023, September 2021. 

\bibitem{sakaguchi_nonlinear_2023}
Atsushi Sakaguchi, Shunya Konno, Fumiya Hanamura,
Warit Asavanant, Kan Takase, Hisashi Ogawa, Petr
Marek, Radim Filip, Jun-ichi Yoshikawa, Elanor Huntington,
Hidehiro Yonezawa, and Akira Furusawa. Nonlinear
feedforward enabling quantum computation. Nature
Communications, 14(1):3817, July 2023.

\bibitem{asavanant_optical_2022}
W. Asavanant and A. Furusawa. Optical Quantum
Computers: A Route to Practical Continuous Variable
Quantum Information Processing. AIP Publishing
Books, 2022.

\bibitem{wolf_motional_2019}
Fabian Wolf, Chunyan Shi, Jan C. Heip, Manuel Gessner,
Luca Pezz`e, Augusto Smerzi, Marius Schulte, Klemens
Hammerer, and Piet O. Schmidt. Motional Fock
states for quantum-enhanced amplitude and phase measurements
with trapped ions. Nature Communications,
10(1):2929, July 2019.

\bibitem{mccormick_quantum-enhanced_2019}
Katherine C. McCormick, Jonas Keller, Shaun C. Burd,
David J. Wineland, Andrew C. Wilson, and Dietrich
Leibfried. Quantum-enhanced sensing of a single-ion
mechanical oscillator. Nature, 572(7767):86–90, August
2019.

\bibitem{wang_heisenberg-limited_2019}
W. Wang, Y. Wu, Y. Ma, W. Cai, L. Hu, X. Mu, Y. Xu,
Zi-Jie Chen, H. Wang, Y. P. Song, H. Yuan, C.-L. Zou,
L.-M. Duan, and L. Sun. Heisenberg-limited single-mode
quantum metrology in a superconducting circuit. Nature
Communications, 10(1):4382, September 2019.

\bibitem{podhora_quantum_2022}
L. Podhora, L. Lachman, T. Pham, A. Le\v{s}und\'{a}k, O. \v{C} \'{ı}p,
L. Slodi\v{c}ka, and R. Filip. Quantum Non-Gaussianity of
Multiphonon States of a Single Atom. Physical Review
Letters, 129(1):013602, June 2022.

\bibitem{deng_heisenberg-limited_2023}
Xiaowei Deng, Sai Li, Zi-Jie Chen, Zhongchu Ni, Yanyan
Cai, Jiasheng Mai, Libo Zhang, Pan Zheng, Haifeng
Yu, Chang-Ling Zou, Song Liu, Fei Yan, Yuan Xu, and
Dapeng Yu. Heisenberg-limited quantum metrology using
100-photon Fock states, June 2023.

\bibitem{pan_realisation_2024}
Xiaozhou Pan, Tanjung Krisnanda, Andrea Duina,
Kimin Park, Pengtao Song, Clara Yun Fontaine, Adrian
Copetudo, Radim Filip, and Yvonne Y. Gao. Realisation
of versatile and effective quantum metrology using
a single bosonic mode, March 2024.

\bibitem{fluhmann_encoding_2019} 
C. Fl\" uhmann, T. L. Nguyen, M. Marinelli, V. Negnevitsky,
K. Mehta, and J. P. Home. Encoding a
qubit in a trapped-ion mechanical oscillator. Nature,
566(7745):513–517, February 2019.

\bibitem{campagne-ibarcq_quantum_2020} 
P. Campagne-Ibarcq, A. Eickbusch, S. Touzard, E. Zalys-
Geller, N. E. Frattini, V. V. Sivak, P. Reinhold, S. Puri,
S. Shankar, R. J. Schoelkopf, L. Frunzio, M. Mirrahimi,
and M. H. Devoret. Quantum error correction of a
qubit encoded in grid states of an oscillator. Nature,
584(7821):368–372, August 2020.

\bibitem{eickbusch_fast_2022}
Alec Eickbusch, Volodymyr Sivak, Andy Z. Ding, Salvatore
S. Elder, Shantanu R. Jha, Jayameenakshi Venkatraman,
Baptiste Royer, S. M. Girvin, Robert J. Schoelkopf,
and Michel H. Devoret. Fast universal control of an oscillator
with weak dispersive coupling to a qubit. Nature
Physics, 18(12):1464–1469, December 2022.

\bibitem{de_neeve_error_2022}
Brennan De Neeve, Thanh-Long Nguyen, Tanja Behrle,
and Jonathan P. Home. Error correction of a logical
grid state qubit by dissipative pumping. Nature Physics,
18(3):296–300, March 2022.

\bibitem{ni_beating_2023}
Zhongchu Ni, Sai Li, Xiaowei Deng, Yanyan Cai, Libo
Zhang, Weiting Wang, Zhen-Biao Yang, Haifeng Yu, Fei
Yan, Song Liu, Chang-Ling Zou, Luyan Sun, Shi-Biao
Zheng, Yuan Xu, and Dapeng Yu. Beating the breakeven
point with a discrete-variable-encoded logical qubit.
Nature, 616(7955):56–60, April 2023.

\bibitem{sivak_real-time_2023}
V. V. Sivak, A. Eickbusch, B. Royer, S. Singh, I. Tsioutsios,
S. Ganjam, A. Miano, B. L. Brock, A. Z. Ding,
L. Frunzio, S. M. Girvin, R. J. Schoelkopf, and M. H. Devoret.
Real-time quantum error correction beyond breakeven.
Nature, 616(7955):50–55, April 2023.

\bibitem{yokoyama_ultra-large-scale_2013}
Shota Yokoyama, Ryuji Ukai, Seiji C. Armstrong,
Chanond Sornphiphatphong, Toshiyuki Kaji, Shigenari
Suzuki, Jun-ichi Yoshikawa, Hidehiro Yonezawa, Nicolas
C. Menicucci, and Akira Furusawa. Ultra-large-scale
continuous-variable cluster states multiplexed in the time
domain. Nature Photonics, 7(12):982–986, December
2013.

\bibitem{arrazola_quantum_2021}
J. M. Arrazola, V. Bergholm, K. Br\' adler, T. R. Bromley,
M. J. Collins, I. Dhand, A. Fumagalli, T. Gerrits,
A. Goussev, L. G. Helt, J. Hundal, T. Isacsson, R. B.
Israel, J. Izaac, S. Jahangiri, R. Janik, N. Killoran, S. P.
Kumar, J. Lavoie, A. E. Lita, D. H. Mahler, M. Menotti,
B. Morrison, S. W. Nam, L. Neuhaus, H. Y. Qi, N. Quesada,
A. Repingon, K. K. Sabapathy, M. Schuld, D. Su,
J. Swinarton, A. Sz\' ava, K. Tan, P. Tan, V. D. Vaidya,
Z. Vernon, Z. Zabaneh, and Y. Zhang. Quantum circuits
with many photons on a programmable nanophotonic
chip. Nature, 591(7848):54–60, March 2021.

\bibitem{enomoto_programmable_2021}
Yutaro Enomoto, Kazuma Yonezu, Yosuke Mitsuhashi,
Kan Takase, and Shuntaro Takeda. Programmable
and sequential Gaussian gates in a
loop-based single-mode photonic quantum processor.
Science Advances, 7(46):eabj6624, 2021.

\bibitem{tiedau_statistical_2021}
J. Tiedau, M. Engelkemeier, B. Brecht, J. Sperling,
and C. Silberhorn. Statistical Benchmarking of Scalable
Photonic Quantum Systems. Physical Review Letters,
126(2):023601, January 2021.

\bibitem{ganapathy_sensing_2023} 
D. Ganapathy, W. Jia, M. Nakano, V. Xu, N. Aritomi, T. Cullen, N. Kijbunchoo, S.E. Dwyer, A. Mullavey et al. (LIGO O4 Detector Collaboration), Broadband Quantum Enhancement of the LIGO Detectors with Frequency-Dependent Squeezing, Phys. Rev. X 13, 041021 (2023)

\bibitem{xia_sensing_2023}
Yi Xia, Aman R. Agrawal, Christian M. Pluchar, Anthony J. Brady, Zhen Liu, Quntao Zhuang, Dalziel J. Wilson, Zheshen Zhang, Entanglement-enhanced optomechanical sensing,
Nature Photonics volume 17, pages 470–477 (2023)

\bibitem{stokowski_sensing_2023}
Hubert S. Stokowski, Timothy P. McKenna, Taewon Park, Alexander Y. Hwang, Devin J. Dean, Oguz Tolga Celik, Vahid Ansari, Martin M. Fejer, Amir H. Safavi-Naeini, Integrated quantum optical phase sensor in thin film lithium niobate,
Nature Communications volume 14, Article number: 3355 (2023) 

\bibitem{guo_sensing_2020}
Distributed quantum sensing in a continuous-variable entangled network
Xueshi Guo, Casper R. Breum, Johannes Borregaard, Shuro Izumi, Mikkel V. Larsen, Tobias Gehring, Matthias Christandl, Jonas S. Neergaard-Nielsen, Ulrik L. Andersen 
Nature Physics volume 16, pages281–284 (2020)

\bibitem{nielsen_sensing_2023}
J.A.H. Nielsen, J.S. Neergaard-Nielsen, T. Gehring, and U.L. Andersen, Deterministic Quantum Phase Estimation  N00N States, Phys. Rev. Lett. 130, 123603 (2023)



\bibitem{chen_experimental_2014}
Moran Chen, Nicolas C. Menicucci, and Olivier Pfister.
Experimental Realization of Multipartite Entanglement
of 60 Modes of a Quantum Optical Frequency Comb.
Physical Review Letters, 112(12):120505, March 2014.

\bibitem{cai_multimode_2017}
Y. Cai, J. Roslund, G. Ferrini, F. Arzani, X. Xu,
C. Fabre, and N. Treps. Multimode entanglement in reconfigurable
graph states using optical frequency combs.
Nature Communications, 8(1):15645, June 2017.


\bibitem{yoshikawa_invited_2016}
Jun-ichi Yoshikawa, Shota Yokoyama, Toshiyuki Kaji,
Chanond Sornphiphatphong, Yu Shiozawa, Kenzo
Makino, and Akira Furusawa. Invited Article: Generation
of one-million-mode continuous-variable cluster state
by unlimited time-domain multiplexing. APL Photonics,
1(6):060801, September 2016.

\bibitem{fisher_mathematical_1920}
R. A. Fisher. A mathematical Examination of
the Methods of determining the Accuracy of Observation
by the Mean Error, and by the Mean
Square Error. Monthly Notices of the Royal
Astronomical Society, 80(8):758–770, June 1920.

\bibitem{knee_fisher_2015}
George C. Knee and William J. Munro. Fisher information
versus signal-to-noise ratio for a split detector.
Physical Review A, 92(1):012130, July 2015.

\bibitem{shannon_mathematical_nodate}
Claude Shannon and Warren Weaver. The Mathematical
Theory of Communication. University of Illinois Press, 1949. 

\bibitem{bowen_tel_2003}
Warwick P. Bowen, Nicolas Treps, Ben C. Buchler, Roman Schnabel, Timothy C. Ralph, Thomas Symul, and
Ping K. Lam. Unity Gain and Nonunity Gain Quantum Teleportation. IEEE Journal of Selected Topics In Quantum Electronics 9, 1519 (2003)

\bibitem{fedorov_tel_2021}
Kirill G. Fedorov, Michael Renger, Stefan Pogorzalek,
Roberto Di Candia, Qiming Chen, Yuki Nojiri
Kunihiro Inomata, Yasunobu Nakamura, Matti Partanen, Achim Marx, Rudolf Gross, Frank Deppe, Experimental quantum teleportation of propagating microwaves, Science Advances 7, sciadv.abk0891 (2021)


\bibitem{andersen_experimental_2005}
Ulrik L. Andersen, Radim Filip, Jarom\'{ı}r Fiur\'{ a}\v{s} ek, Vincent Josse, and Gerd Leuchs. Experimental purification of coherent states. Physical Review A, 72(6):060301, December 2005.

\bibitem{lassen_experimental_2010}
Mikael Lassen, Lars Skovgaard Madsen, Metin Sabuncu,
Radim Filip, and Ulrik L. Andersen. Experimental
demonstration of squeezed-state quantum averaging.
Physical Review A, 82(2):021801, August 2010.

\bibitem{yonezu_time-domain_2023}
Kazuma Yonezu, Yutaro Enomoto, Takato Yoshida,
and Shuntaro Takeda. Time-Domain Universal Linear-
Optical Operations for Universal Quantum Information
Processing. Physical Review Letters, 131(4):040601, July
2023. 

\bibitem{takeda_-demand_2019}
Shuntaro Takeda, Kan Takase, and Akira Furusawa.
On-demand photonic entanglement synthesizer. Science
Advances, 5(5):eaaw4530, May 2019.



\bibitem{scully_quantum_1997}
Marlan O. Scully and M. Suhail Zubairy. Quantum
optics. Cambridge University Press, Cambridge, 1997.
	
\bibitem{bachor_guide_2009}	
Hans-Albert Bachor and Timothy C. Ralph. A guide to
experiments in quantum optics. Wiley-
VCH, Weinheim, 2009.

\bibitem{glorieux_hot_2023}
Quentin Glorieux, Tangui Aladjidi, Paul D Lett, and
Robin Kaiser. Hot atomic vapors for nonlinear and quantum
optics. New Journal of Physics, 25(5):051201, May
2023.

\bibitem{vaidya_broadband_2020}
V. D. Vaidya, B. Morrison, L. G. Helt, R. Shahrokhshahi,
D. H. Mahler, M. J. Collins, K. Tan, J. Lavoie,
A. Repingon, M. Menotti, N. Quesada, R. C. Pooser,
A. E. Lita, T. Gerrits, S. W. Nam, and Z. Vernon.
Broadband quadrature-squeezed vacuum and nonclassical
photon number correlations from a nanophotonic device.
Science Advances, 6(39):eaba9186, September 2020.

\bibitem{zhang_squeezed_2021}
Y. Zhang, M. Menotti, K. Tan, V. D. Vaidya, D. H.
Mahler, L. G. Helt, L. Zatti, M. Liscidini, B. Morrison,
and Z. Vernon. Squeezed light from a nanophotonic
molecule. Nature Communications, 12(1):2233, April
2021.

\bibitem{cheng_bivariate_2020}
Cheng, Weijun and Xiaoting Wang.
“Bivariate Fisher–Snedecor F Distribution and its Applications in Wireless Communication Systems.”
IEEE Access 8 (2020): 146342-146360.

\bibitem{dong_experimental_2008}
R. Dong, M. Lassen, J. Heersink, C. Marquardt, R. Filip, G. Leuchs and U. L. Andersen, Experimental entanglement distillation of mesoscopic quantum states, Nature Physics 4 (2008), 919-923

\bibitem{dong_continuous_2010}
R. Dong, M. Lassen, J. Heersink, C. Marquardt, R. Filip, G. Leuchs and U. L. Andersen, Continuous-variable entanglement distillation of non-Gaussian mixed states, Phys. Rev. A 82, 012312 (2010)

\bibitem{ruppert_fading_2019}
L. Ruppert, C. Peuntinger, B. Heim, K. Gunthner, V. C. Usenko, D. Elser, G. Leuchs, R. Filip and C. Marquardt, Fading channel estimation for free-space continuous-variable secure quantum communication, New Journal of Physics 21, 123036 (2019)

\bibitem{usenko_entanglement_2012}
V. C. Usenko, B. Heim, C. Peuntinger, C. Wittmann, C. Marquardt, G. Leuchs and R. Filip, Entanglement of Gaussian states and the applicability to quantum key distribution over fading channels, New Journal of Physics 14, 093048 (2012) 

\bibitem{fiurasek_conditional_2001}
J. Fiurášek, Conditional generation of sub-Poissonian light from two-mode squeezed vacuum via balanced homodyne detection, Phys. Rev. A 64, 053817 (2001)

\bibitem{schnabel_squeezed_2017}
R. Schnabel, Squeezed states of light and their applications in laser interferometers, Physics Reports 684, 1-51 (2017)

\bibitem{glockl_squeezed-state_2006}
O. Gl\" ockl, U. L. Andersen, R. Filip, W. P. Bowen,
and G. Leuchs. Squeezed-State Purification with Linear
Optics and Feedforward. Physical Review Letters,
97(5):053601, August 2006.

\bibitem{van_loock_detecting_2003}
Peter Van Loock and Akira Furusawa. Detecting genuine
multipartite continuous-variable entanglement. Physical Review A, 67(5):052315, May 2003.

\bibitem{gu_quantum_2009}
Mile Gu, Christian Weedbrook, Nicolas C. Menicucci,
Timothy C. Ralph, and Peter van Loock. Quantum
computing with continuous-variable clusters. Physical
Review A, 79(6):062318, June 2009.

\bibitem{gonzalez-arciniegas_cluster_2021}
Carlos Gonz\' alez-Arciniegas, Paulo Nussenzveig, Marcelo
Martinelli, and Olivier Pfister. Cluster States
from Gaussian States: Essential Diagnostic Tools for Continuous-Variable One-Way Quantum Computing.
PRX Quantum, 2(3):030343, September 2021.

\bibitem{zapletal_experimental_2021}
Petr Zapletal, Tom Darras, Hanna Le Jeannic, Adrien
Cavaill\' es, Giovanni Guccione, Julien Laurat, and Radim
Filip. Experimental Fock-state bunching capability of
non-ideal single-photon states. Optica, 8(5):743, May
2021.



	
\end{thebibliography}
\end{document}